\documentclass[acmsmall]{acmart}

\AtBeginDocument{%
  \providecommand\BibTeX{{%
    \normalfont B\kern-0.5em{\scshape i\kern-0.25em b}\kern-0.8em\TeX}}}

\newcommand{\modif}[1]{{#1}}

\newcommand{\tech}[0]{{{\sc Technique}}}
\newcommand{\task}[0]{{{\sc TaskType}}}

\newcommand{\wordname}[0]{WordTouch}
\newcommand{\chunkname}[0]{ChunkTouch}
\newcommand{\baseline}[0]{{Baseline}}

\newcommand{\onedim}[0]{WordTouch}
\newcommand{\semantic}[0]{ChunkTouch}

\newcommand{\word}[0]{{{\it Word}}}
\newcommand{\phrase}[0]{{{\it Phrase}}}
\newcommand{\nonphrase}[0]{{{\it Non-phrase}}}
\newcommand{\clause}[0]{{{\it Clause}}}
\newcommand{\halfSent}[0]{{{\it Half-sentence}}}
\newcommand{\sentence}[0]{{{\it Sentence}}}
\newcommand{\twosent}[0]{{{\it Two-sentences}}}
\newcommand{\wtend}[0]{{{\it Word-to-end}}}
\newcommand{\wparag}[0]{{{\it Whole-text}}}

\newcommand{\sunit}[0]{{{\it Semantic Units}}}
\newcommand{\nsunit}[0]{{{\it Non-semantic Units}}}

\newcommand{\sysname}[0]{{{1D-Touch}}}

\newcommand{\tct}[0]{{\small {\it TimeCompletion}}} 

\newcommand{\oversh}[0]{{\small {\it NumOvershoots}}}
\newcommand{\attempt}[0]{{\small {\it NumAttempts}}}

\newcommand{\anovase}[4]{{($F_{#1, #2} = #3$, $p < .01$, \eff{#4})}}
\newcommand{\anovasse}[4]{{($F_{#1, #2} = #3$, $p < .001$, \eff{#4})}} 
\newcommand{\eff}[1]{$\eta_p^2 = #1$}

\newcommand{\mean}[1]{$M = #1$}

\newcommand{\wttest}[2]{{(\mean{#1}, $W = #2$, $p < .05$)}}
\newcommand{\wttests}[2]{{(\mean{#1}, $W = #2$, $p < .01$)}}

\newcommand{\twottest}[6]{{($M_{#1} = #2$, $M_{#3} = #4$, $T_{#5} = #6$, $p < .05$)}}

\setcopyright{rightsretained}
\acmPrice{}
\acmDOI{10.1145/3626483}
\acmYear{2023}
\copyrightyear{2023}
\acmSubmissionID{iss23main-p1771-p}
\acmJournal{PACMHCI}
\acmVolume{7}
\acmNumber{ISS}
\acmArticle{447}
\acmMonth{12}
\received{2023-07-01}
\received[accepted]{2023-09-22}

\begin{document}


\title{1D-Touch: NLP-Assisted Coarse Text Selection via a Semi-Direct Gesture}

\author{Peiling Jiang}
\email{peiling@ucsd.edu}
\orcid{0000-0003-4447-0111}
\authornote{This work was done when the author was an intern at the City University of Hong Kong.}
\affiliation{%
  \department{School of Creative Media}
  \institution{City University of Hong Kong}
  \city{Hong Kong}
  \country{China}
}
\affiliation{%
  \institution{University of California San Diego}
  \streetaddress{9500 Gilman Dr}
  \city{La Jolla}
  \state{California}
  \country{USA}
  \postcode{92093}
}

\author{Li Feng}
\email{Lfeng256@connect.hkust-gz.edu.cn}
\orcid{0000-0002-6198-0896}
\affiliation{%
  \department{School of Creative Media}
  \institution{City University of Hong Kong}
  \city{Hong Kong}
  \country{China}
}
\affiliation{%
  \institution{The Hong Kong University of Science and Technology (Guangzhou)}
  \city{Guangzhou}
  \country{China}
}

\author{Fuling Sun}
\email{fulingsun@ucsd.edu}
\orcid{0000-0003-0453-0757}
\affiliation{%
  \institution{University of California San Diego}
  \streetaddress{9500 Gilman Dr}
  \city{La Jolla}
  \state{California}
  \country{USA}
  \postcode{92093}
}

\author{Parakrant Sarkar}
\email{parakrantsarkar@gmail.com}
\orcid{0000-0001-6505-3443}
\affiliation{%
  \department{School of Creative Media}
  \institution{City University of Hong Kong}
  \city{Hong Kong}
  \country{China}
}

\author{Haijun Xia}
\email{haijunxia@ucsd.edu}
\orcid{0000-0002-9425-0881}
\affiliation{%
  \institution{University of California San Diego}
  \streetaddress{9500 Gilman Dr}
  \city{La Jolla}
  \state{California}
  \country{USA}
  \postcode{92093}
}

\author{Can Liu}
\email{canliu@cityu.edu.hk}
\orcid{0000-0003-3267-3317}
\affiliation{%
\department{School of Creative Media}
  \institution{City University of Hong Kong}
  \city{Hong Kong}
  \country{China}
}
\authornote{Corresponding author.}

\renewcommand{\shortauthors}{Jiang, et al.}

\begin{abstract}
Existing text selection techniques on touchscreen focus on improving the control for moving the carets. Coarse-grained text selection on word and phrase levels has not received much support beyond word-snapping and entity recognition. We introduce 1D-Touch, a novel text selection method that complements the carets-based sub-word selection by facilitating the selection of semantic units of words and above. This method employs a simple vertical slide gesture to expand and contract a selection area from a word. The expansion can be by words or by semantic chunks ranging from sub-phrases to sentences. This technique shifts the concept of text selection, from defining a range by locating the first and last words, towards a dynamic process of expanding and contracting a textual semantic entity. To understand the effects of our approach, we prototyped and tested two variants: WordTouch, which offers a straightforward word-by-word expansion, and ChunkTouch, which leverages NLP to chunk text into syntactic units, allowing the selection to grow by semantically meaningful units in response to the sliding gesture. Our evaluation, focused on the coarse-grained selection tasks handled by 1D-Touch, shows a 20\% improvement over the default word-snapping selection method on Android.
\end{abstract}



\begin{teaserfigure}
  \includegraphics[width=\textwidth]{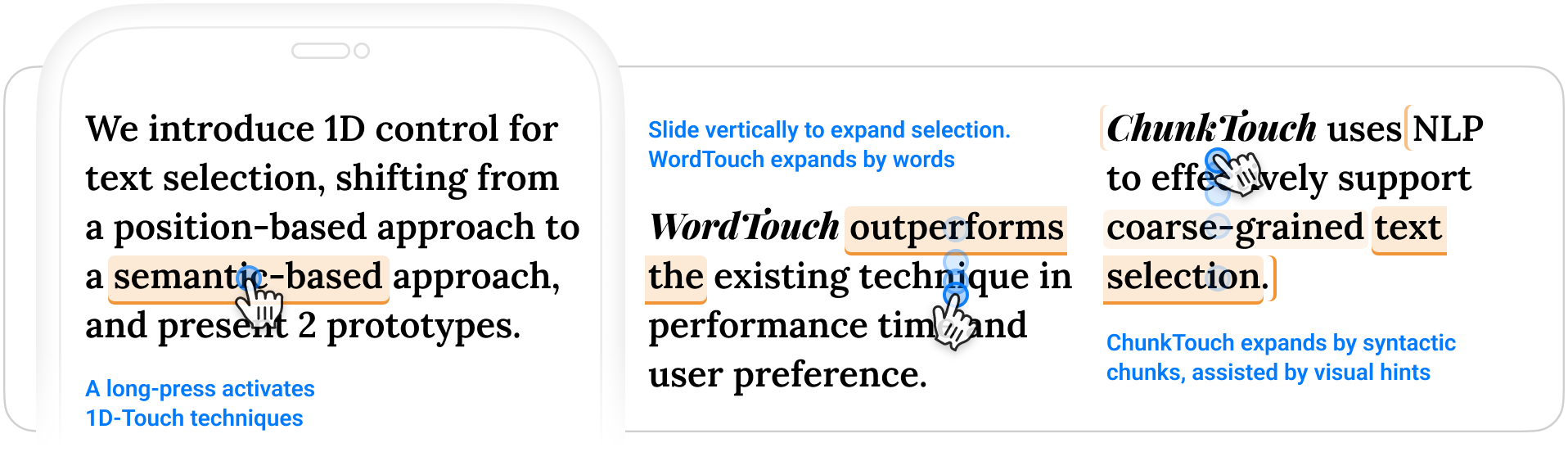}
  \caption{The illustrated interface of two \sysname{} techniques prototypes---\wordname{} and \chunkname{}.}
  \Description{The illustrated interface of two \sysname{} techniques prototypes---\wordname{} and \chunkname{}.}
  \label{fig:teaser}
\end{teaserfigure}

\maketitle

\section{Introduction}

Text selection through direct manipulation on touchscreens is known to be difficult due to the Fat Finger Problem \cite{siek2005fat}. The target acquisition task itself consists of defining a range between the start and end positions. The fundamental interface used for this task involves moving a pair of carets to these positions between words or characters. To mitigate the challenge of direct manipulation with low precision input, previous research focused on introducing \emph{indirection} to text selection by providing indirect control of caret positions \cite{Magnifier-widget,Ando2018,Ando2019}.

Although character-level text selection is important, there are many text manipulation tasks that focus on a larger granularity of text that conveys semantic meanings with words, phrases, clauses, or sentences. Typical examples include highlighting key phrases for note-taking, selecting phrases to be translated, copying-pasting Named Entities (NE) or other semantic segments, etc. Furthermore, there are increasing needs and opportunities for supporting semantic-based text manipulation, with the rapid rise of speech-based multimodal text manipulation, which often operates on the word-level~\cite{10.1145/3490099.3511103}, and the breakthroughs in Large Language Models (LLMs) that operates on the tokens (i.e. words).

Previous research showed that providing word-snapping for manipulating carets can improve the performance of some text selection tasks~\cite{MIURA20141644}. In fact, both Android and iOS have integrated a word-snapping technique in their text selection method, in which users begin with a long-press (or double-tap) on a word to select it, and subsequently, without lifting the finger, drag it to the target\footnote{\url{https://support.apple.com/guide/iphone/select-and-edit-text-iph1a9cae52c/ios}}. The selection snaps to the end of the word closest to the touch lift position. This technique is integrated seamlessly with the carets for character-level selection by having the handles shown after the word-level selection is finished as the finger first lifts.

Beyond word-snapping, semantic-based text selection includes selecting units of phrase, sentence, and paragraph~\cite{Goguey2018, LEE2020} as well as ``smart text selection'' assisted by Natural Language Processing (NLP) that recognizes user intent~\cite{pantel-etal-2014-smart} or common entities in the text ~\cite{bier2006entity}, such as addresses, URLs, and phone numbers\footnote{\url{https://android-developers.googleblog.com/2017/05/whats-new-in-android-o-developer.html}}. Previous research also introduced the concept of ``fuzzy text selection'' and demonstrated how it can be well-used for text highlighting for sense-making tasks on mobile \cite{chang2016supporting} and for assisting active reading with multimodal input \cite{hinckley2012informal}.

Building on existing work, this paper introduces \sysname{}---a new approach for coarse-grained text selection in the granularity of \emph{words and above}. Using one-dimensional input, \sysname{} shifts the concept of text selection, from specifying two 2D positions for defining a range, to selecting one word as the initial target and gradually expanding it to adjacent words to form a syntactic or semantic chunk of selection.
Our technique can also be set to expand the selection by semantic units in sizes of words, sub-phrases, clauses, and sentences.
We prototyped \sysname{} by leveraging the constituency parsing in existing NLP tools to chunk text, providing a more gradual and advanced segmentation than basic units that only include words, phrases, sentences, and paragraphs.

\modif{The novelty of our approach is twofold: the use of a semi-direct one-dimensional sliding gesture for text selection; and the continuous and multi-scale linguistic chunking of selection units.}
To test the effects of the gestural control and the semantic chunking respectively, we conducted a controlled experiment that compared variants of \sysname{} prototypes with expansion by words (\wordname{}) and expansion by chunks (\chunkname{}) with the state-of-the-art selection technique on Android with word-snapping (\baseline{}). Our results showed that both of our \sysname{} techniques outperformed the baseline significantly in overall performance time by 20\%, with \chunkname{} being particularly good for selecting semantically meaningful units.

The three main contributions of this paper include 1) a new approach for text selection that combines direct and indirect manipulation gestures; 2) two \sysname{} techniques that prototype the concept and enable evaluation of factors contributing to the advanced selection performance; and 3) empirical findings that demonstrate significant performance gain from both the simple semi-direct control and the NLP-assisted text chunking method, as well as provide insights into future improvements and integration with character-level selection methods.

\section{Related Work}

This section summarizes existing text selection techniques, many of which have been employed to mitigate the difficulty of controlling carets on touchscreens, such as adding a magnifier \cite{Magnifier-widget}, using the keyboard as trackpad \cite{Ando2019}, or tilting the device to control selection \cite{Ando2018}.
We summarize these efforts in providing \emph{indirect} or \emph{direct} control for caret positioning, as well as other novel approaches leveraging \emph{alternative modalities}, such as gaze or touch pressure to extend the users' degrees of freedom of control.

\subsection{Indirect Control}

Applying mapping strategies to enable users to control the handle positions without directly touching them is one remedy to the fat finger problem. 
The keyboard, as a familiar tool in text editing scenarios, is usually leveraged as the virtual controller \cite{lai2019shortcut, Fuccella2013, Ando2019, Ando2018, zhang2019gedit}. \modif{Fuccella et al. presented a series of gesture controls for text selection, where the selection range can be extended by swiping two fingers on the keyboard left or right \cite{Fuccella2013}. This method expands the selection by one word per swipe, which is designed for fine-tuning an existing selection during text editing rather than selecting longer targets across multiple sentences.}
Ando et al. pegged particular keys to navigate the cursor, leaving other keys for extending the selection range \cite{Ando2019}. Press \& Tilt controls the selection by tilting the phone while pressing a key \cite{Ando2018}. \modif{However, unconventional tilting gestures require additional practice, which makes them more challenging than conventional touch- and handle-based techniques.}
Gedit proposes two types of gestures to control the cursor, including ring and flick, which were proven to be easy to learn and conducive for efficient revision for selected text \cite{zhang2019gedit}.
ForceSelect allows users to adjust the text selection granularity on mobile headsets by controlling the force they apply to press on the virtual trackpad with a mapping strategy \cite{LEE2020}.
Suzuki et al. investigated switching the frame of reference, enabling users to move the rest of the interface around while keeping the caret fixed at the middle of the screen \cite{suzuki2016fix}. \modif{Evaluations demonstrated significant improvements in selection time compared to the default UI, however, only for very small text (12 pt).}

\subsection{Direct Control}

A few techniques have been proposed to address the Fat Finger Problem while maintaining direct control over the caret handles.
Smart Selection leverages linguistic analysis through an ensemble of learning approaches to determine the users' intentions based on their touchpoints \cite{pantel-etal-2014-smart}. \emph{Multi-step} approaches have been leveraged to select precisely with touch --- the user begins by providing a rough localization of the target text, after which they can adjust the precision of the selection with a zoom-in interface \cite{Roudaut2008, chen2014bezelcopy, li2020swap}. \modif{BezelCopy uses bezel gestures, whereby the sentence is selected by dragging from the left bezel to the sentence on the screen \cite{chen2014bezelcopy}. It then creates a new fullscreen page highlighting only the selected sentence with enlarged font and handle sizes, allowing the user to perform the conventional handle-based selection and copy with more confidence. Complete copy-paste workflows were evaluated, and the results showed that the technique outperforms the system-default pipeline for common tasks.
Esteves et al. explored the use of touch-based motion matching to indirectly select out-of-reach targets~\cite{Esteves2022One}.}
Swap by Li et al. is useful for text selection during revision by swapping individual words within a given text \cite{li2020swap}. After typing in a new word at the end of the text, the user taps the area around the word to be changed, which converts texts in its vicinity into enlarged buttons for easy selection. This eliminates the need to focus on the tiny carets during text selection. Additionally, Goguey et al. introduced a force-sensitive selection technique that leveraged a virtual ``mode gauge'' that displays the touch pressure when performing selection \cite{Goguey2018}. Different pressure levels are mapped to different levels of chunking, and the gauge visualizes the pressure changes.

\subsection{Alternative Modalities}

While our technique focuses exclusively on touch input, we have also reviewed a few prior works that rely on alternative or multiple modalities to assist in text selection. Several techniques adopt touch pressure and force to provide an additional dimension of input control \cite{Goguey2018, LEE2020, Lee2021}. However, while empirical results of these efforts showed benefits for some selection targets, they did not exhibit an overall significant performance gain in terms of selection time and error rate, compared to the standard caret selection \cite{LEE2020}.

\modif{Gaze is also leveraged as additional information to assist text selection and editing \cite{Rivu2020, Biedert2010}. However, empirical results also question the effectiveness, e.g., Gaze'N'Touch performs better than the touch-based technique when it comes to larger texts but not to a significant degree \cite{Rivu2020}. More recently, EyeSayCorrect leverages users' gaze trajectory to implicitly select a word to be edited with voice, which saves text selection time by 40\%~\cite{10.1145/3490099.3511103}. However, this technique is dedicated to error correction and can select only one word at a time. Gaze-Shifting, on the other hand, uses gaze to modulate input and combines direct pointing (e.g., through a touch pen) and indirect mapping through the user's gaze that offsets the input to the visual target~\cite{Pfeuffer2015gaze}.}

\subsection{Novelty of Our Approach}

Prior research has investigated various ways to improve text selection, yet most of them focused on manipulating carets, either directly or indirectly. Despite the Fat Finger Problem persists in the standard caret selection technique, new interventions demonstrate limited overall performance gain. Our work takes a different path and focuses on introducing a semi-direct control to improve text selection at the word level and above.

There are two main novelties in this approach. First, indirect input and user control are introduced differently in the technique. Instead of indirectly controlling the position of the cursor or carets, the technique employs a gesture that starts with direct manipulation (touch) on a word and ends with an \emph{indirect mapping} between distance increment and the number of included semantic entities in the target. Second, we introduce a continuous and generic NLP-assisted text selection method that snaps the ends of the selection to a syntactic or semantic unit, which differs from related works that spot only special entities in the text, such as addresses or phone numbers.

\section{\sysname{} Techniques}

Text represents a one-dimensional stream of information that is placed sequentially into a two-dimensional form for printing or displaying on screens. Existing text selection techniques echo this dimensional dissonance as they accomplish the target acquisition by finding the positions of the two ends.
Our work, on the other hand, introduces the use of one-dimensional control for text selection with \textbf{\sysname{}}, a new technique that allows users to select text with a vertical 1D sliding gesture. The user slides up and down to expand or contract the selected range.
The expansion and contraction of the selection range are controlled by the sliding distances\footnote{An online demo of the \sysname{} techniques is deployed at \url{https://text-selection.netlify.app}. Please access with a device with a touch screen.}.

How can we segment text into distinct semantic units that represent chunks of meaning? In cognitive science literature, such ``meaning'' is often described as the ``gist'' extracted from a given verbatim, functioning at varying degrees of abstraction~\cite{REYNA19951}. Consequently, we define semantic units of text as chunks that can vary in size and represent meanings at different levels of abstraction. As an initial exploration, we choose \emph{word} as the smallest \emph{semantic} unit, with the \emph{syntactic} unit serving as a method to chunk text that mirrors its underlying semantics.

We built two variants of the technique that adopt the 1D control method---\textbf{\wordname{}} and \textbf{\chunkname{}}.
\wordname{} employs a simple word-by-word expansion controlled by the sliding distance, while \chunkname{} expands by syntactic units that grow larger in chunks derived from NLP Constituency Parsing. This section introduces how we designed and implemented these two \sysname{} techniques in detail.

\begin{figure*}[t]
  \centering
  \includegraphics[width=0.95\textwidth]{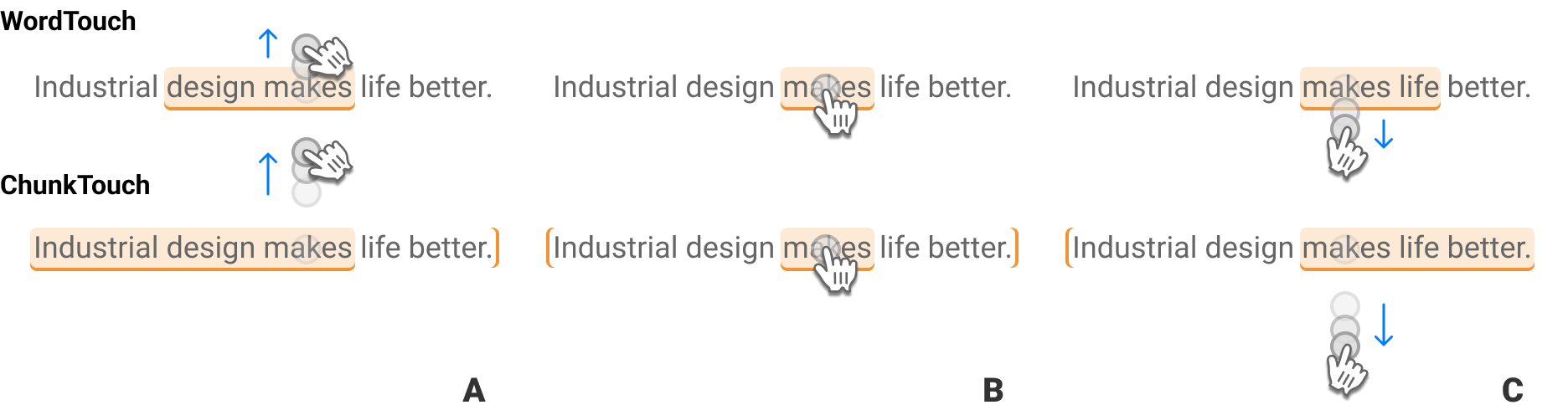}
  \caption{\sysname{} selects by sliding up or down. Activated by a conventional long-press, the word under the touch point gets selected as the initial state (B). When sliding up, the selection expands by words (\wordname{}) or syntactic chunks (\chunkname{}) to the text before (A); and vice versa for sliding down (C).}
  \Description{\sysname{} selects by sliding up or down. Activated by a conventional long-press, the word under the touch point gets selected as the initial state (B). When sliding up, the selection expands by words (\wordname{}) or syntactic chunks (\chunkname{}) to the text before (A); and vice versa for sliding down (C).}
  \label{fig:directions}
\end{figure*}

\subsection{Selecting by 1D Sliding} 

To start the text selection, the user activates the \sysname{} techniques by directly long-pressing on a word, highlighting its background. We use the conventional 500 ms threshold for detecting the long-press which is the same as the standard selection techniques on Android and iOS.
Once activated, the user can slide up or down to expand the selection from the first selected word to the words before or after it, as shown in Figure~\ref{fig:directions}.
The selection keeps growing as the user continues sliding.
Haptic feedback is provided for activation as well as every unitary expansion, similar to the default vibration feedback for long-press on Android and iOS devices.

\subsubsection{Mapping the Sliding Distance to the Number of Expansion Units}

We map the sliding distance on the y-axis (as compared to the initial touchpoint, in millimeters) to the number of words (for \wordname{}) or syntactic chunks (for \chunkname{}) to be added to the selection.

\begin{equation}
\label{eq:number}
    N = \lfloor \frac{25.4}{\mathrm{PPI}} \cdot \frac{p}{d} \rfloor
\end{equation}

\begin{figure}
  \centering
  \includegraphics[width=0.45\linewidth]{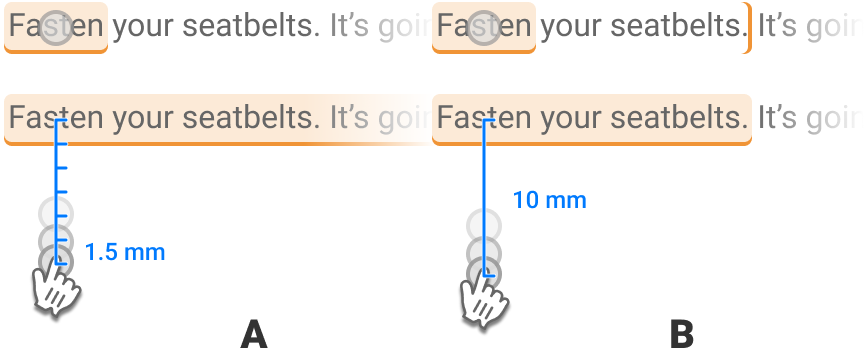}
  \caption{\wordname{} (A) expands to include the next word when sliding every 1.5 mm, and \chunkname{} (B) merges the next chunk when sliding every 10 mm.}
  \Description{\wordname{} (A) expands to include the next word when sliding every 1.5 mm, and \chunkname{} (B) merges the next chunk when sliding every 10 mm.}
  \label{fig:trigger}
\end{figure}

Equation~\ref{eq:number} describes how we determine the number of words or chunks ($N$) to expand, where $\mathrm{PPI}$ is the pixel density of the screen, $p$ is the vertical sliding distance in pixels, and $d$ is the predefined triggering distance of expansion for each technique. Based on our own experiences in selecting with \sysname{} techniques on the device used for the following experiments (with a 6.7-inch screen), we set $d_\mathrm{\onedim{}} = 1.5$ and $d_\mathrm{\semantic{}} = 10$ for our implementation (Figure~\ref{fig:trigger}). We noticed that \chunkname{} expansion needed a longer triggering distance because users needed more time to confirm the next chunk of text to be selected. Shorter triggering distances caused more overshoots. Thus we chose 10 mm as a trade-off between selection efficiency and overshoot frequency.

More research needs to be conducted to optimize these thresholds in the future. They can also be adaptive for different screen sizes and user preferences (possibly through a calibration process to be discussed in Section~\ref{sec:future}).

\subsubsection{Rewinding the Overshot Selections}
Overshoot is a common problem in text selection where one selects more than they originally planned to select.
\sysname{} techniques support rewinding to fix an overshoot. After overshooting, before lifting the finger, the user can slide backward to rewind the current selection. The reference point for rewinding is the furthest point they have previously reached, so that the rewinding slide can go beyond the initial touchpoint. Rewinding is always by words, and we use the same mapping relationship as \wordname{} for both techniques. As a result, when using \chunkname{}, if the expanded chunk is longer than the text the user intended to select, they can use rewinding as a remedy. \chunkname{} thus supports selecting \emph{from any word to any word} in a corpus without sacrificing the efficiency of expanding the selection.
This differs from other chunking methods in existing works that are limited to selecting up to five fixed levels of granularity---word, phrase, sentence, paragraph, and the whole text \cite{MIURA20141644, li2020swap, Goguey2018}.

\subsubsection{Clutching with Increasing Target Sizes}
\sysname{} supports clutching, which allows users to continue expanding an existing selection by sliding again.
If one shot of sliding does not expand the selection enough to include all target text, users can lift the finger and start another sliding gesture from \emph{anywhere} on the selected text. 
As a result, the selection becomes increasingly easier when the user clutches, as the target area grows bigger. While clutching, the selection always starts with expansion towards the sliding direction, while the user can retract by sliding backward without lifting the finger.
Because our technique supports expanding to both directions (up and down), enabling clutching to directly retract would conflict with expanding to the other direction. After expanding, the user could then retract any number of words, even reducing the selection area to be smaller than before the clutching was performed, to correct overshoot from a previous gesture.

\subsection{Chunking Text by Syntactic Units}

\begin{figure}[t]
    \centering
    \begin{minipage}[t]{.49\textwidth}
        \centering
        \includegraphics[width=\linewidth]{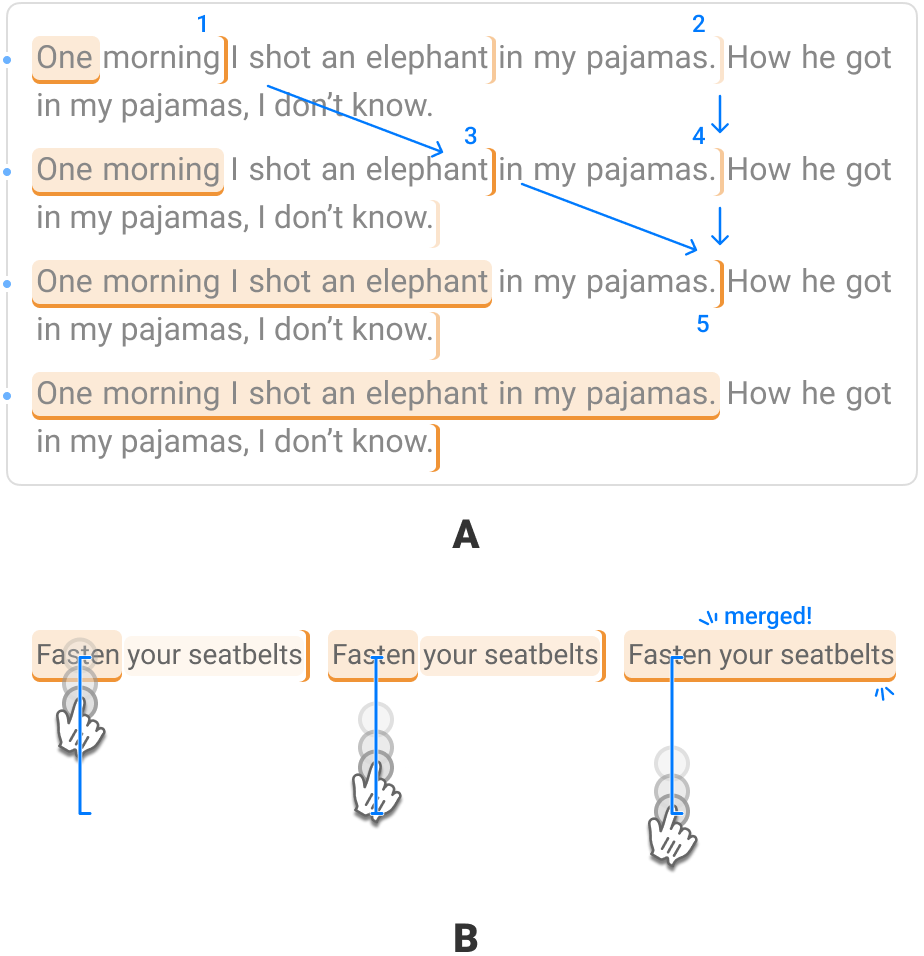}
        \caption{(A) The brackets shift as the selection changes. (B) Changing backgrounds visualize sliding progress.}
        \Description{(A) The brackets shift as the selection changes. (B) Changing backgrounds visualize sliding progress.}
        \label{fig:branckets}
    \end{minipage}
    \hfill
    \begin{minipage}[t]{.49\textwidth}
        \centering
        \includegraphics[width=\linewidth]{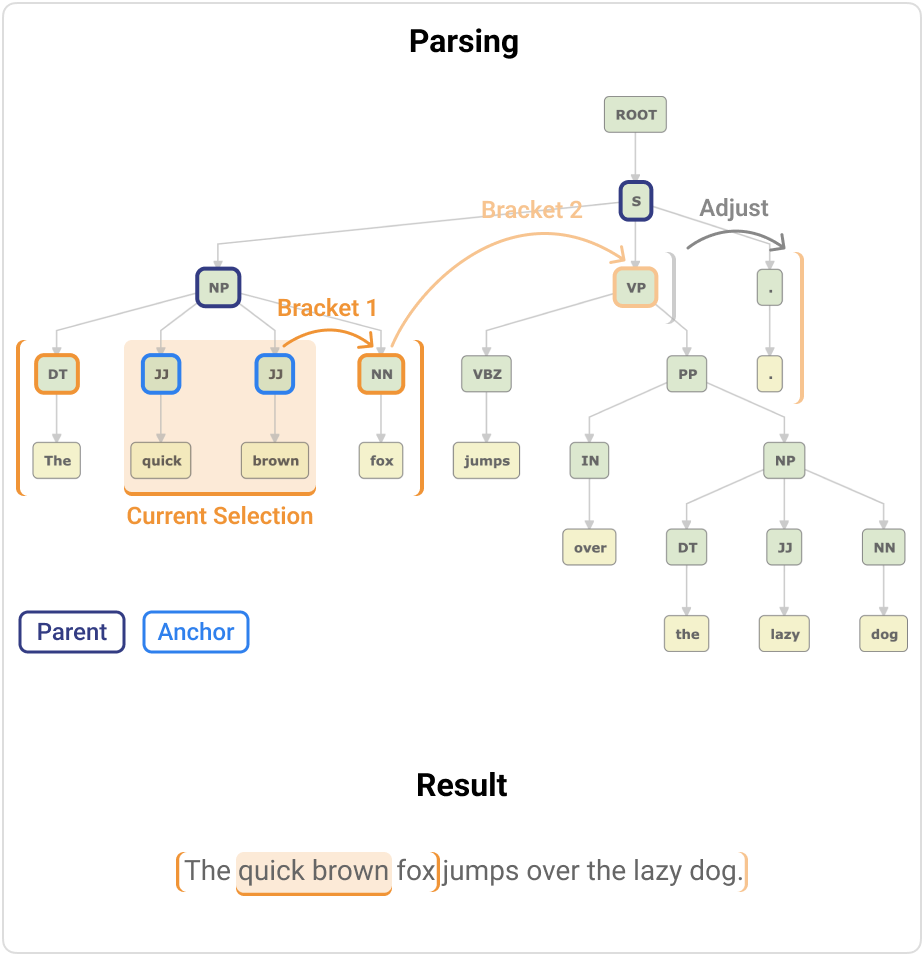}
        \caption{An example NLP Constituency Parsing process and its result, as visualized by the brackets.}
        \Description{An example NLP Constituency Parsing process and its result, as visualized by the brackets.}
        \label{fig:nextSibling}
    \end{minipage}
\end{figure}

We prototyped \chunkname{} by chunking text in syntactic units, as a first-step approach to infer units that represent meaning. As shown by the example in Figure~\ref{fig:branckets}A, the next text chunks to expand can be of any length of phrases, clauses, sentences, etc. We discuss how we identify these syntactic chunks in Section~\ref{sec:algorithm}. As the lengths of the expanding chunks vary, which increases the uncertainty for selection actions, we design the visual feedforward and feedback mechanisms to assist using the \chunkname{} technique.

\subsubsection{Visual Feedforward and Feedback Design} Extra feedforward and feedback visualizations are provided for \chunkname{} to facilitate the gestural control and text selection.

\paragraph{Brackets show future chunks based on the current selection} We introduce \emph{brackets}, placed before and after the current selection, as a visual feedforward to inform users of the future chunks that will be merged into the current selection. The brackets that are most visible and closest to the current selection indicate the endpoints of the next chunk to be merged, as in Figure~\ref{fig:branckets}A.1. The lighter brackets provide a preview of the next level of expansion. We display three sets of brackets that expand from the selected text in this implementation. As selection grows, the adjacent brackets move simultaneously, e.g., in Figure~\ref{fig:branckets}A, the closest bracket moves from position 1 to 3, and then to 5. Through a different perspective, users can also see the opacity of the brackets change, as in Figure~\ref{fig:branckets}A.2, 4, and 5, signifying the approaching expansion to the end of this sentence. For \chunkname{}, we actively compute brackets for every change in the selection and always display them with the selection.

\paragraph{Changing the background highlights the immediate chunk to expand} As the user starts sliding, we color the background of the to-be-merged chunk to semi-transparent (Figure~\ref{fig:branckets}B). As the sliding distance increases, the closer the expansion is to be triggered, the less transparent the background of the text chunk will be, \modif{up to sixty percent of the background density of the selected text that creates a discrete visual indication when a chunk becomes actually selected.} The semi-transparent background works as feedback to actively inform the user of the upcoming chunk and visualize their sliding progress. Together with feedforwarding brackets, the changing-color background makes the action status clearer and increases the controllability of \chunkname{}.

\subsubsection{Segmenting Text with NLP}
\label{sec:algorithm}

To segment text and identify neighboring syntactic chunks, we leverage an existing constituency parser from CoreNLP, a state-of-the-art NLP model to perform syntactic analysis on the corpus \cite{zhu2013fast, manning2014stanford}. The corpus will first be pre-processed into a tree-like structure that can be used for the following selection activities (Figure~\ref{fig:nextSibling}). The CoreNLP parser divides the corpus by sentences into disconnected trees and assigns a ROOT for each of them. We then added one more whole-text-level ROOT as the parent of them to build a tree that connects all the sentences. We refer to the green part-of-speech (POS) tags for the level of each word.
\modif{Although the parsing quality can be affected by the syntactic and grammatical accuracy of the original text, we observed that the CoreNLP parser remains robust across various texts and consistently delivers tree structures in a uniform format. The parser demonstrates sufficient robustness for handling in-the-wild user-provided text with typos and grammar errors.}

We determine the next syntactic chunks by the user's current selection---we anchor the first and last word of the selected text to find the adjacent chunks, namely \emph{siblings}, that are positioned before and after the current selection.
We identify the sibling chunks to be the adjacent syntactic nodes at the same or closest possible higher levels in the constituency tree.
This strategy balances the expansion efficiency and granularity with a presumption that users prefer to expand at the same level of parsing as they have initially selected, e.g., expanding by phrases when a phrase is selected, and expanding by sentences when a sentence is selected. As a result, the more one selects, the faster \chunkname{} expands with larger chunks.

More specifically, we look up the next sibling by finding the node that is adjacent to and shares the same parent with the anchoring node. If none were found, e.g., node \emph{NN--fox} in Figure~\ref{fig:nextSibling} has no adjacent node after it, we recursively set the parent of the current anchoring node as the new anchor, and search for its siblings, until reaching the whole-text ROOT, where we know that we have selected the first or last word of the whole corpus.
Once an immediate sibling has been found, we place brackets at the respective position or adjust it so that the neighboring punctuation would be included (only for the chunks after). Then, we leverage the end of this chunk as the new anchor, and recursively repeat the steps above, until reaching the corpus ends or the maximum number of brackets needed (3 in the current implementation).

\subsection{Implementing a Working Prototype}

We implement our technique as a web application with React. Each word and punctuation mark is wrapped in an \emph{SemanticNode} element. A \emph{SemanticBlock} contains multiple nodes and represents the current segmentation, which can be merged with other blocks or further divided into smaller units. A selected block renders the selection range in colors. A paragraph is represented by a \emph{SemanticText} element and included by a \emph{SemanticBody} element with other paragraphs. Touch-event listeners are bound with each paragraph node to detect selection events.

For \chunkname{}, we use Node.js \cite{dahl-2021} to run a wrapped version of CoreNLP~\cite{bort-2019}, originally only available in Python. \emph{SemanticBody} detects text changes and sends the new text, divided into paragraphs, to our server for parsing. The results are usually returned within only a fraction of a second and are then processed by our segmentation algorithm and filters. When waiting for the parsing result, or when an error occurs in parsing the text, our interface places every node at the same depth as the instant child of the sentence ROOT as a fallback---\chunkname{} then expands by words (adjacent siblings) just like \wordname{}.

\section{Study}

To evaluate the \sysname{} approach, we conducted a controlled experiment to compare its efficiency and accuracy against a baseline technique for \emph{word-level} selection. We compared \onedim{} and \semantic{} in separate experimental conditions in order to distinguish the effects from the 1D control and the semantic chunking. Based on our experience with the technique, the following hypotheses are made:

\begin{itemize}
    \item H1: Both \semantic{} and \onedim{} outperforms \baseline{} in overall performance time.
    \item H2: \semantic{} performs better than \onedim{} in selecting \sunit{}, while worse in selecting \nsunit{}.
\end{itemize}

\subsection{Baseline}

The baseline technique we chose is the default selection technique provided in the Android system. \modif{As mentioned in the Introduction section, it has a word-snapping feature activated by a 500 ms long-press on a word as the start word. Without lifting the finger, the user drags it towards the end of the selection. The system actively snaps the end of the selection to the end of the word closest to the touch position.} While the selection expands by words, its endpoint moves backward character-by-character when users retract the selection to allow finer adjustment. This word-snapping feature is also integrated with the caret by having the handles shown after users first release their fingers. Users can then further adjust the selection by character with the handles. 
To our knowledge, this is still the state-of-the-art text selection technique, which is shown to perform better or comparably well in recent literature \cite{Goguey2018, Darbar2021}. We refer to this technique as \baseline{} in this section.

\subsection{Experiment Design}
\label{sec:exp_design}

\begin{figure}[t]
    \centering
    \begin{minipage}[t]{.49\textwidth}
        \centering
        \includegraphics[width=\linewidth]{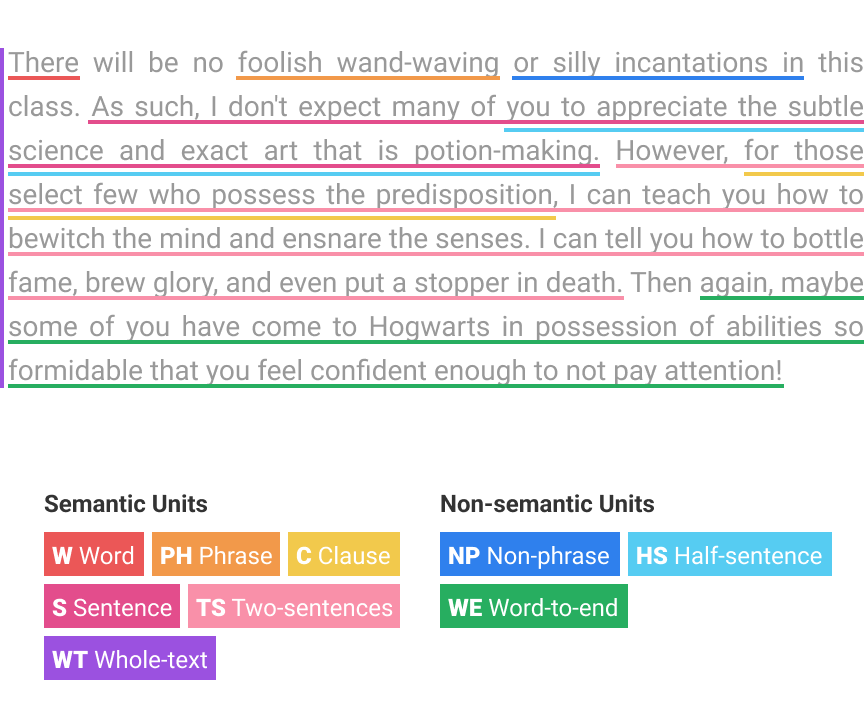}
        \caption{Examples of a target piece of text for each \task, for semantic units and non-semantic units.}
        \Description{Examples of a target piece of text for each \task, for semantic units and non-semantic units.}
        \label{fig:task_examples}
    \end{minipage}
    \hfill
    \begin{minipage}[t]{.49\textwidth}
        \centering
        \includegraphics[width=\linewidth]{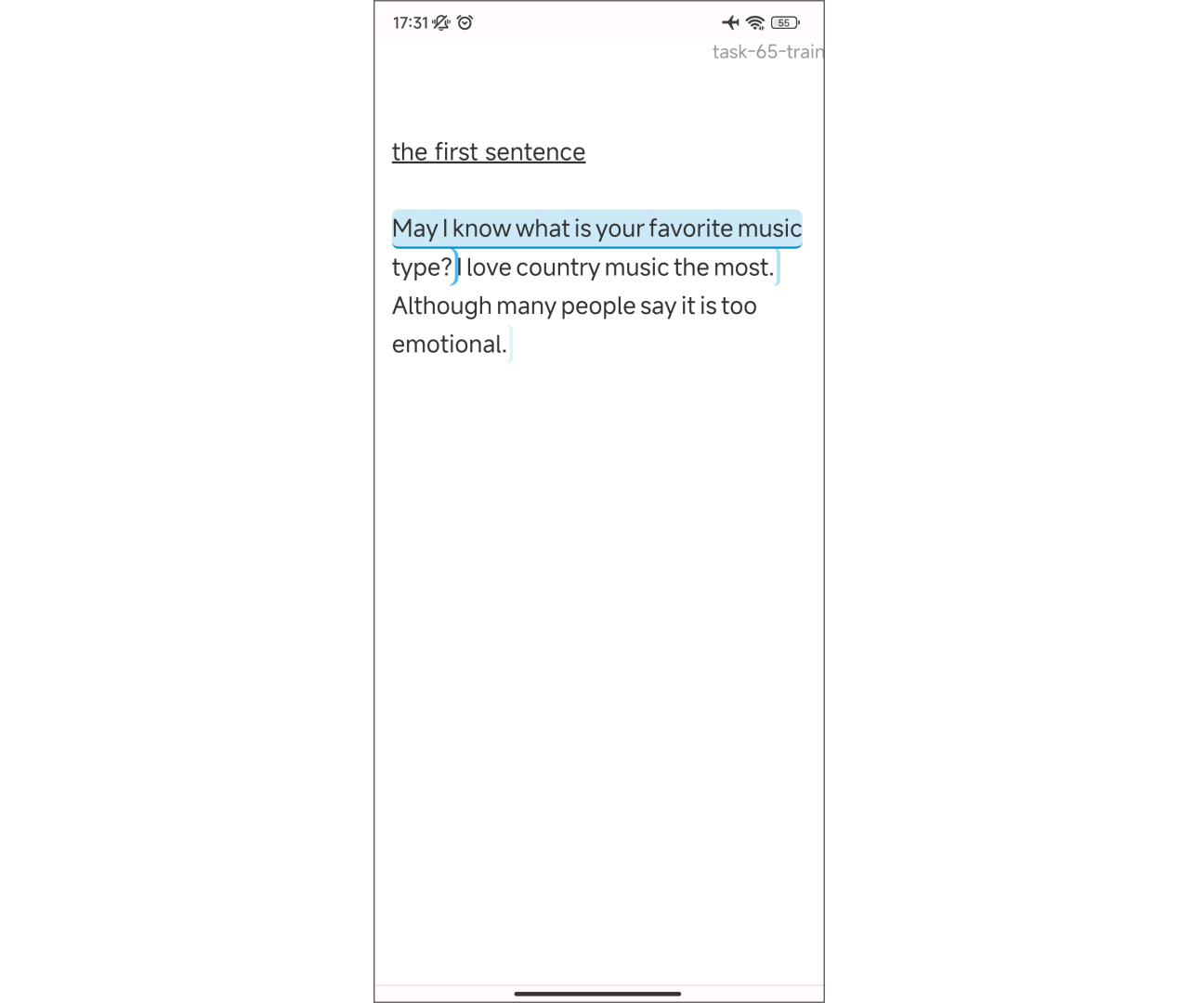}
        \caption{The screenshot of an example task in the study in the \semantic{} condition.}
        \Description{The screenshot of an example task in the study in the \semantic{} condition.}
        \label{fig:study}
    \end{minipage}
\end{figure}

Our $3 \times 9$ within-subject experiment had two factors: \tech{} [\semantic{}, \onedim{}, \baseline{}], and \task{} [\word{}, \phrase{}, \nonphrase{}, \clause{}, \halfSent{}, \sentence{}, \twosent{}, \wtend{}, and \wparag{}].

The 9 task types are applied to a collection of textual excerpts that are different in type and length, as well as their positions within the given corpus.
Following prior research, these tasks were chosen as they cover a variety of selections that can be required in real-world use~\cite{Goguey2018, Darbar2021}.
Instead of applying the original nine types of tasks like the prior works, we removed character-selection tasks and replaced character-to-end tasks with word-to-end tasks, as our technique is designed to support word level and above granularity of selection.

The selection tasks differ in two main parameters. One is the length: \word{} $<$ \phrase{} and \nonphrase{} ($\approx$ four words) $<$ \clause{} and \halfSent{} ($\approx$ ten words) $<$ \sentence{} $<$ \twosent{} $<$ \wtend{} $<$ \wparag{}.
The other parameter is whether the selected text forms a semantically meaningful unit or not, which we will refer to as either \sunit{} or \nsunit{} herein.
We created three additional types of tasks with selection targets that do not form semantic units: \nonphrase{} in a similar length to \phrase{}, \halfSent{} in a similar length to \clause{}, and \wtend{}. Since it is likely that whether the selection target is a semantic unit or not would affect the performance of \semantic{}, during the result analysis, we investigated this as an independent variable in order to better understand the effect and the usability of \semantic{} for the tasks that both align and in-align with the design of the technique.

Our technique leverages the same activation method as the baseline and can potentially be integrated with the caret-based character-level adjustment in the same way as well: carets show after the user finishes the sliding gesture and lifts the finger. Although the integration deserves additional testing, we chose to focus this experiment on testing the tasks that depend only on the novel aspects of \sysname{} and avoid introducing additional factors by reverse-engineering the system-default carets interface, which is required for prototyping the integration.

\subsection{Tasks and Instructions}
For each trial, we display the target text at the top of the screen (Figure~\ref{fig:study}).
We create the selection tasks based on the parameters mentioned above. 9 \task{} $\times$ 4 repetitions $=$ 36 unique tasks are created for each \tech{}, resulting in 108 unique tasks in total.
We composed all these texts to maintain similar lengths and structures for the same \task{} and to ensure the syntactic units are easily distinguishable. The difficulty for comprehension is kept as easy to understand for high school students or above.
To avoid any noise introduced by parser errors from CoreNLP, we tested the corpus to ensure the syntactic segmentation returned in runtime was free of error.

\subsection{Participants}
We recruited twelve undergraduate students (five male, seven female) from local universities as participants. None of the participants had knowledge about our techniques before.

\subsection{Apparatus}

We built a web application that depended on a local server (for CoreNLP, experiment monitoring, and data collection) to run the experiment. Due to technical reasons, the first half of the study was conducted on a Samsung Galaxy S20+ (6.7 inch, 3200 $\times$ 1440 px) for participants to complete the tasks with a server running on a Windows desktop. The other half of the studies ran with a Google Pixel 7 (6.3 inch, 2400 $\times$ 1080 px) and a Mac server. In both cases, the server and client communicate through Socket.io. The implementation of \semantic{} uses Node.js to run a wrapped version of CoreNLP (version 4.5.4) on the server side. No significant difference in results was identified for these two setups.

Instructions and text to be selected were displayed in typical size, font (Roboto Regular), and layout to facilitate reading (Figure~\ref{fig:study}). The font size is adjusted so that the lowercase `x' is displayed in 2 mm in height on the screen. The left-aligned target text had a line height of $160\%$, with side margins of 3 mm from screen edges, and the content was placed 50 mm below the screen top.

\subsection{Procedure}

Participants were first briefed about our technique and the purpose of this experiment. The experiment contained three blocks, one for each \tech{}. The order of techniques was counterbalanced among twelve participants. \emph{Playground} sessions were included at the beginning of all blocks for participants to practice, where no time limit was set. During the experiment, participants generally spent around 2--3 minutes in the playground. At the beginning of each block, there were nine training trials before starting the monitored trials, one for each \task{}. For the monitored tasks, four consecutive trials were repeated for each \task{} whilst the order of \task{} was randomized in each block.
A trial was automatically completed when the selected text matched the expected result and the user lifted the finger.
Finally, participants filled in a questionnaire after finishing all blocks.
An experiment session took about thirty minutes.

\subsection{Data Collection}

Every touch and selection-change event was recorded with a timestamp. We then pre-processed the raw logs and further collected the following measures for each trial: (1) \tct{}, \emph{Task completion time} is defined as the duration from the first TouchDown event to the last SelectionChange event. (2) \oversh{}, \emph{Number of overshoots} is the number of times one selects more than the required part of the text. For subjective experiences and preferences, we collected the perceived task load with the NASA-TLX index \cite{HART1988139} and gathered their preferences by asking them to rank the most and least preferred techniques.
For \sysname{} techniques, we logged TouchDown, TouchMove, TouchEnd, and SelectionChange events. Due to the limited access to the system's default selection techniques, we only logged TouchDown and SelectionChange events for \baseline{}.

\section{Results}

We removed outliers that were potentially caused by accidental touch or interruption. We excluded trials with \tct{} that are more than three standard deviations from the mean \tct{} of the respective technique and task type. We excluded 10 trials from \baseline{} ($2.0\%$), 12 trials from \semantic{} ($2.6\%$), and 13 trials from \onedim{} ($2.8\%$). Due to the violation of the normality assumption, we performed a non-parametric Aligned Rank Transform (ART) for a two-way repeated measures ANOVA ($\alpha = .05$, with Greenhouse-Geisser correction) on \tct{} and \oversh{}. The Wilcoxon signed-rank tests with the Benjamini-Hochberg procedure were performed as the post hoc tests.

\begin{figure}
  \centering
  \includegraphics[width=0.53\linewidth]{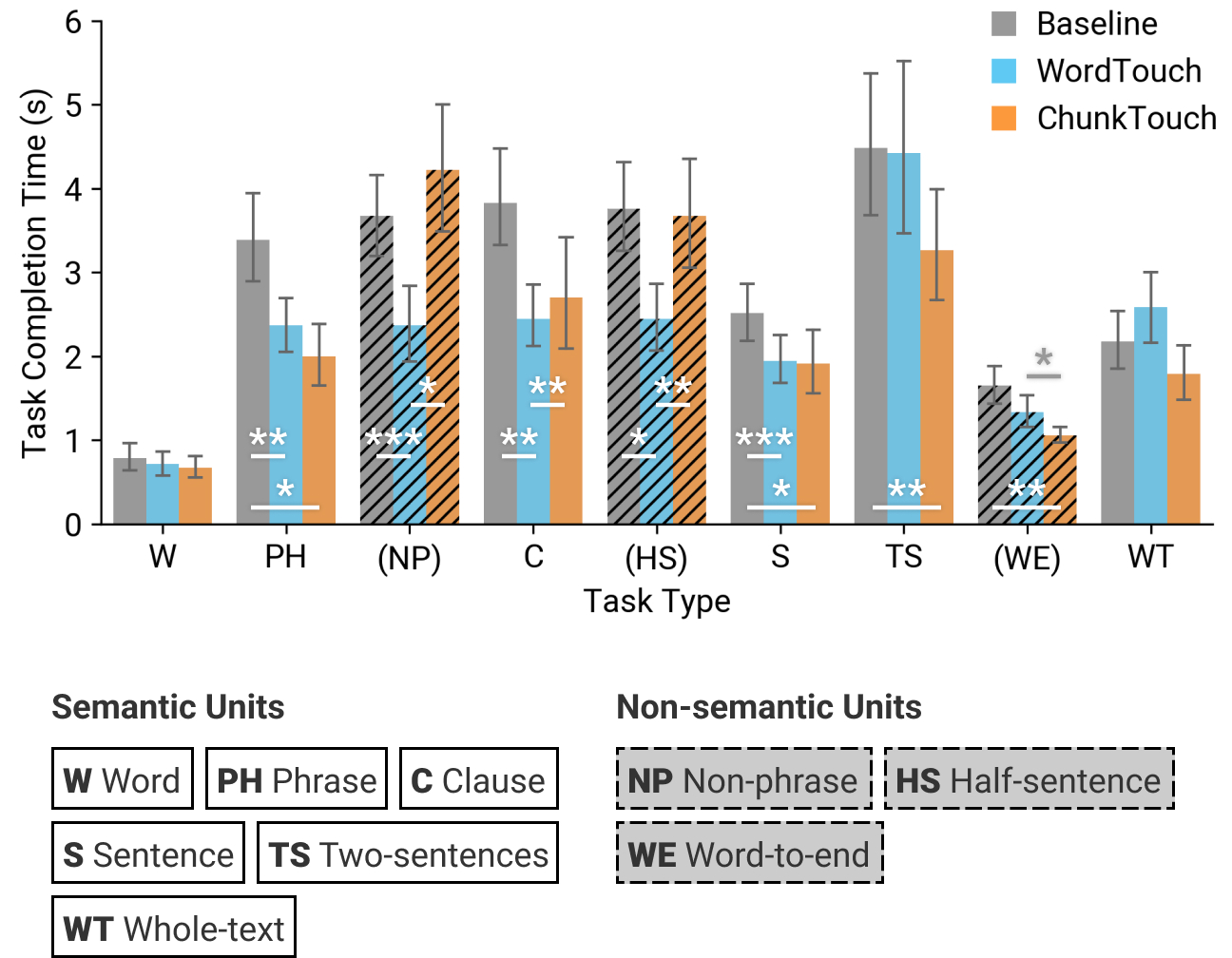}
  \caption{Performance time for different \task{} and \tech{}. Task type labels enclosed in parentheses with shaded bars indicate the target text for the corresponding task type is non-semantic. Significant results are labelled as *: $p < .05$, **: $p < .01$, ***: $p < .001$. The error bars represent the 95\% confidence intervals.}
  \Description{Performance time for different \task{} and \tech{}.}
  \label{fig:results_1}
\end{figure}

\subsection{Performance Time}

The results supported both of our hypotheses regarding performance time.

\subsubsection{H1 is Supported}
\label{sec:h1}

\begin{figure}
    \centering
    \begin{minipage}[t]{.5\textwidth}
        \centering
        \includegraphics[width=.796125\linewidth]{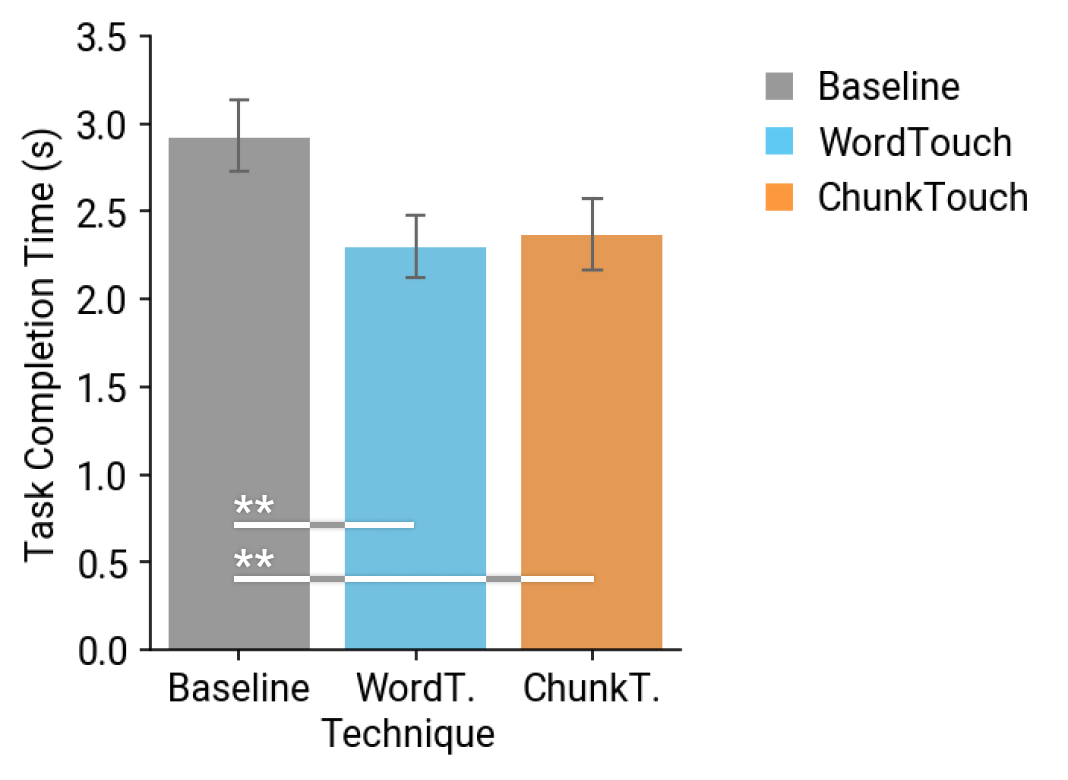}
        \caption{Performance time for each technique.}
        \Description{Overall performance time for each technique.}
        \label{fig:results_2}
    \end{minipage}%
    \begin{minipage}[t]{.5\textwidth}
        \centering
        \includegraphics[width=.95\linewidth]{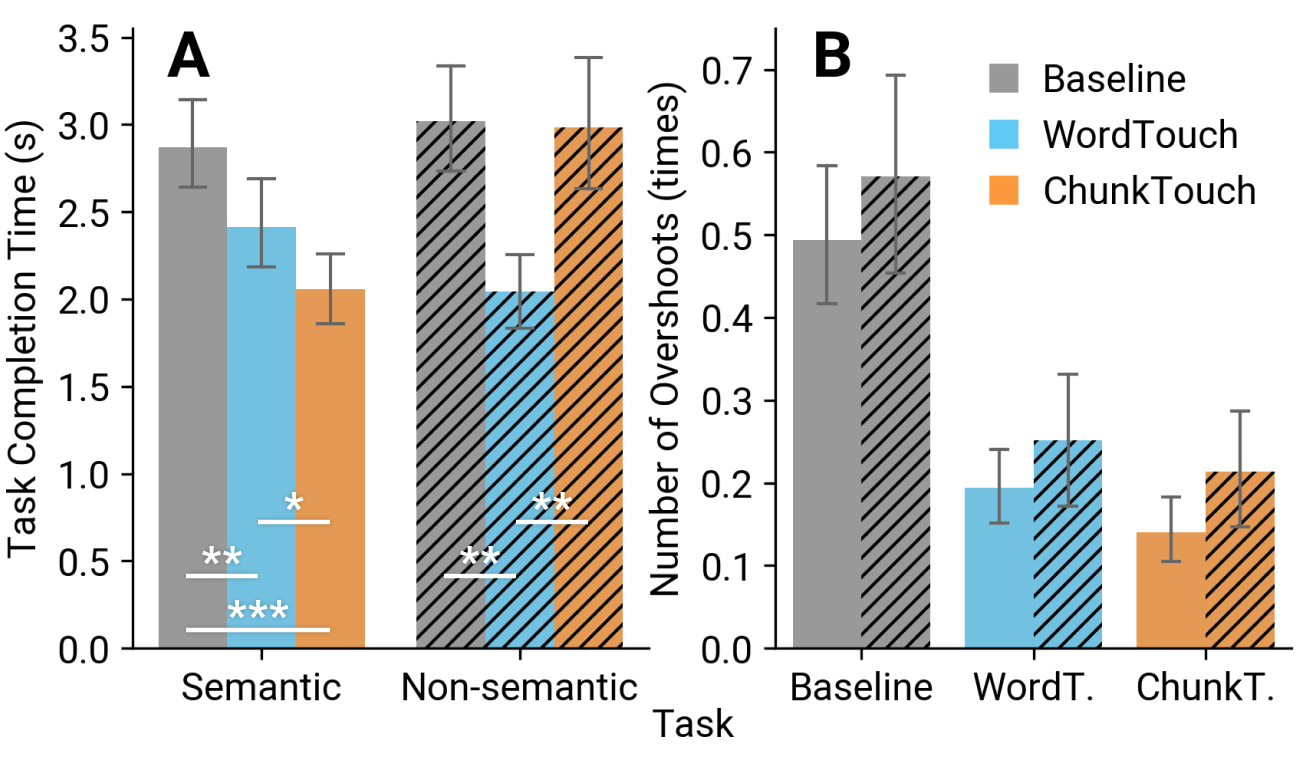}
        \caption{(A) Performance time for each technique, for semantic and non-semantic targets. (B) The number of overshoots for different \tech{}, for semantic (solid) and non-semantic (shaded) targets.}
        \Description{(A) Performance time for each technique, for semantic and non-semantic targets. (B) The number of overshoots for different \tech{}, for semantic (solid) and non-semantic (shaded) targets.}
        \label{fig:results_3}
    \end{minipage}
\end{figure}

Both \onedim{} and \semantic{} perform significantly faster than \baseline{} in the overall \tct{}.
As shown in Figure~\ref{fig:results_2}, the two-way repeated measures ANOVA on ranked \tct{} showed significant effects of \tech{} \anovase{2}{22}{23.614}{.176}, \task{} \anovasse{8}{88}{89.710}{.700}, and their interaction \anovasse{16}{176}{4.764}{.185}.
Post hoc tests showed that, when using \onedim{}, the participants spent significantly less time \wttests{2292 \mathrm{\ ms, \textbf{21\%\ faster\ than\ \baseline{}}}}{0} to finish a trial, compared to when selecting with \baseline{} (\mean{2919 \mathrm{\ ms}}). \semantic{} \wttests{2361 \mathrm{\ ms, \textbf{19\%\ faster\ than\ \baseline{}}}}{3} also significantly outperformed \baseline{}.

As shown in Figure~\ref{fig:results_1}, the post hoc tests showed that the participants selected text significantly faster with \onedim{} than with \baseline{} when selecting \phrase{}, \nonphrase{}, \halfSent{}, and \sentence{}; and performed faster with \semantic{} for \phrase{}, \clause{}, \sentence{}, \twosent{}, and \wparag{}\footnote{Detailed statistical analysis results of post hoc tests can be found in Appendix~\ref{app:test}.}. Meanwhile, we found no significant difference between the three techniques for \word{} and \wparag{} tasks.

\subsubsection{H2 is Supported}

We calculated the performance time for \sunit{} and \nsunit{} tasks by taking the averages of all the tasks listed under each category in Figure~\ref{fig:results_1}.
For \sunit{}, \semantic{} ($M = 2055\mathrm{\ ms}$) was significantly faster than \onedim{} \wttest{2415 \mathrm{\ ms}}{13}, by 15\%. Both \onedim{} and \semantic{} performed significantly faster than \baseline{} ($M_\mathrm{Baseline} = 2868 \mathrm{\ ms}$, \onedim{}: $W = 7$, $p < .05$, 16\% faster; \semantic{}: $W = 2$, $p < .01$, 28\% faster). For \nsunit{}, \semantic{} ($M = 2984\mathrm{\ ms}$) performed significantly worse than \onedim{} \wttests{2045 \mathrm{\ ms}}{0}, 46\% slower (Figure~\ref{fig:results_3}A). \onedim{} still significantly outperformed \baseline{} in this case ($M_\mathrm{Baseline} = 3020 \mathrm{\ ms}$, $W = 3$, $p < .01$).
While \semantic{} was still faster than \baseline{}, no significant difference was found.

More specifically, \semantic{} performed significantly worse than \onedim{} when selecting \nsunit{} including \nonphrase{}, \halfSent{}, and \wtend{}. On the other hand, post hoc tests showed no significant difference between the two \tech{} for other types of tasks.

\subsection{Selection Accuracy}
\label{sec:accuracy}

Another major measure of selection performance is the error rate. Correcting errors in selection takes time and affects user experience. A lower error rate means higher selection accuracy.
We measured the error rate through \attempt{} and \oversh{} for each trial.

\subsubsection{Selection Attempts}

\attempt{} is defined as the number of times one clears all the selections, by tapping outside the selected area, to restart the process.
We found a significant main effect for \task{} on \attempt{}. The post hoc tests showed that only significant pairwise differences existed between \baseline{} and \onedim{} \twottest{\baseline{}}{1.113}{\onedim{}}{1.070}{11}{3.570}, and \semantic{} and \baseline{} \twottest{\semantic{}}{1.066}{\baseline{}}{1.113}{11}{3.483}, for \phrase{} selections. Participants needed to restart significantly more often when using \semantic{} than the other two techniques. We believe this is a main contributor to the time cost in \semantic{} for \phrase{} tasks because of the $500 ms$ activation time for every restart.
 
\subsubsection{Overshoot}
Both \sysname{} techniques and \baseline{} start the selection at the word level, and expand to the text before and after this selection by either sliding up or down or by dragging the carets. Overshoot, i.e. expanding too much that exceeds the target point, is a common selection inaccuracy that 
could lead to longer selection time and more selection attempts. Overshoot could be costly for \sysname{}, especially \semantic{}. For \baseline{}, however, it is not a problem as it is easy to correct. As shown in Figure~\ref{fig:results_3}B, participants overshot significantly more with \baseline{} than our techniques.
We did not find any significant difference between \onedim{} and \semantic{}. There is also no difference between \sunit{}  and \nsunit{} for each technique respectively.
From our observation, since overshooting is more costly in \semantic{}, participants tend to avoid it by reducing the touch speed, opting to undershoot.

\subsection{Subjective User Experience}

We collected the perceived task load with the NASA-TLX index \cite{HART1988139}, asked the participants to rank each technique, and interviewed them about their experience in using three selection techniques. The results are reported below.

\subsubsection{Perceived Workload} According to the collected feedback of a 7-point NASA Task Load Index questionnaire, none of the measured dimensions (Mental demand, Physical demand, Temporal demand, Performance, Effort, and Frustration) showed significant differences between \baseline{}, \onedim{}, and \semantic{}.

\subsubsection{User Preference}
\begin{figure}[t]
  \centering
  \includegraphics[width=0.5\linewidth]{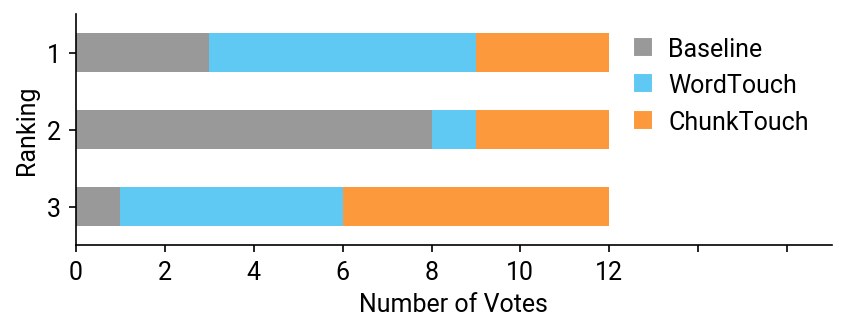}
  \caption{Participant ranking of three techniques.}
  \Description{Participant ranking of three techniques.}
  \label{fig:results_ranking}
\end{figure}

As shown in Figure~\ref{fig:results_ranking}, 9 out of 12 participants ranked one of the \sysname{} techniques---\onedim{}---as the most preferred technique (the first rank).
While the participants found \baseline{} familiar (P0: \textit{``already used to it''} and P11: \textit{``can go letter by letter as I am used to''}), most of them preferred \onedim{} as it is more efficient, intuitive, and easy to control (P2: \textit{``direct and fast''} and P8: \textit{``intuitive to interact''}). For \semantic{}, some participants agreed that chunking helped them select \sunit{} and longer text (P5: \textit{``phrases and sentences were easy to select''}). Some appreciated the visual feedback (P4: \textit{``the various levels of color and brackets are informative''}). Despite the preferences, a chi-squared test on vote results did not show significant differences among the three techniques.

Participants also mentioned frustrations with all techniques. For \baseline{}, it was harder to select longer pieces of text (P0, P7, P10), required extra attention to deal with characters and punctuation (P1 and P4), and the tiny carets were hard to grab (P2 and P8). For \onedim{}, it is harder to rewind (P0, P2, P5, and P7) and the selection could be occluded by the finger body (P4). For \semantic{}, some participants (P3--P5) found it hard to rewind due to the reduced rewind speed (i.e. expanding by syntactic chunks but rewinding by words). And P3 was confused by the visualization ( \textit{``The color changed. So I would think I was selecting the sentences afterward.''}). Constructive suggestions were also brought up, including adding a zoom-in view as a remedy for occlusion (like the traditional handle-based techniques) and making the \textit{sensitivity} customizable. The issues and suggestions, which will be further discussed in Section~\ref{sec:future}, will assist us in developing future iterations of \sysname{} techniques and in exploring meaning-driven text selection.

\subsubsection{Selection Strategies}

We observed several strategies that participants leveraged when using our techniques. P5 found brackets in \chunkname{} helpful in focusing on the adjacent text that is about to be selected (\textit{``as it made me look at the chunk rather than the whole text''}).
P6 found himself more careful when using our techniques due to the cost of rewinding (\textit{``\semantic{} required me to be mindful about where I put my finger after my first selection. When using \onedim{}, I struggled to make edits to my initial selection so I ended up just canceling the entire selection.''}). However, P0 drew the opposite conclusion---\textit{``I normally use carets by first deciding a range and then the details, but \semantic{} can help me determine the range.''}

\subsubsection{Semantic Chunking}

Most participants found the chunking of \semantic{} ``makes sense'' (P6), especially in selecting longer text (P6: \textit{``chunking more flexibly than only by sentences or paragraphs makes a lot of sense''} and P11: \textit{``easier to select long targets with chunking''}). P6 found the chunking helped him select faster (\textit{``Having the system recognize semantically-related words for quicker selection of smaller chunks of text.''}). P3 also reported the \emph{inefficiency} of \semantic{} in selecting non-semantic units (\textit{``If you're trying to get parts of a phrase, I didn't like how you had to unselect after selecting something.''}) The comments indicate that the subjective experience with the interface is consistent with our quantitative findings that \semantic{} performs better in selecting semantic units while having trade-offs in selecting non-semantic units.

\subsubsection{Visual and Haptic Feedback}
\label{visualFeedbak}

The participants appreciated the various feedback provided in our techniques, including the brackets and color changes for \semantic{}, as well as haptic feedback for both techniques. Many participants appreciated the visual design (P7: \textit{``color changing assisted me in making selection more accurately''}). While P3 found it a bit confusing and suggested \textit{``changing the color altogether rather than just having a lighter color''}. In addition, P10 found the brackets confusing (\textit{``I'm always confused by where the bracket is.''}). P6 suggested making the \textit{``anchor points''} (i.e. brackets) more salient.

Most participants find the haptic feedback for both \onedim{} and \semantic{} helpful (P1: \textit{``The vibration was helpful, as you can feel what you have selected even before seeing the visual feedback.''}). P6 preferred haptic feedback and thought it improved the experience of \emph{indirect} control (\textit{``Since my finger isn't directly on top of what I'm selecting, it's nice to have some haptic feedback, other than visual feedback, to tell me I'm doing something.''}).

\subsubsection{User Acceptance and Use Cases}
We asked participants if they wanted to use \onedim{} or \semantic{} for daily text selection. 8 out of 12 participants wished to use \onedim{} on a daily basis (P5: \textit{``selecting some words when sending emails to others''} and P11: \textit{``\onedim{} is probably best for smaller selections where selecting like one word at a time is useful.''}). 7 participants claimed that they wished to employ \semantic{} in various daily activities, especially when selecting longer text (P0: \textit{``quoting some longer sentences''} and P6: \textit{``working with a lot of text for copy and paste''}). While \semantic{} has a learning curve, most participants believed it would become more convenient to use after sufficient practice.

\section{Discussion}
\label{sec:future}

In this work, we introduce the \sysname{} approach, which uses one-dimensional semi-direct sliding gestures for text selection in the granularity of word level and above. Our study showed that our approach outperformed the start of the art in efficiency and on par with it in accuracy.
Syntactic chunking segments the text into units, from words to larger entities parsed through NLP. While our choices of control (sliding distance) and chunking method (constituency parsing) can find alternatives in the future, evaluating these two example techniques---\wordname{} and \chunkname{}---allowed us to gain insights into the \sysname{} approach.

\subsection{Text Selection as Linguistic Object Expansion}
The fundamental conceptual shift \sysname{} brings is that word-level text selection can be less about finding the positions of the first and last words (or characters) on the screen, and more about growing the selected text \emph{object} with syntactically or semantically associated units.
This may be counter-intuitive for beginners as most people are used to position-based text selection, while easy to get used to based on our experience and user study observations. Multiple clutching also gets increasingly easier with each consecutive stroke as the expanded text increases the target area for the finger to touch.
\sysname{}'s control also makes it potentially more suitable for selecting longer text where the target ends off-screen on a scrollable interface. With the baseline, the off-screen selection is time-based---users have to hold the handle at the edge of the screen and wait for the page to roll up while aiming to stop at a good position.

\subsection{Semi-Direct Input with Discrete CD Gain}
A unique characteristic of this technique is that a selection starts with a \emph{direct} touch on the first word and ends at a relative position based on \emph{indirect} control. It is known in the literature that both direct and indirect input methods have unique advantages~\cite{hinckley2007input}. For example, direct manipulation allows users to input naturally and intuitively, while indirect input enables manipulating out-of-reach targets with enhanced precision. Our method allows an intuitive direct selection of the starting point, while the endpoint is specified indirectly by growing the target, thus mitigating the Fat Finger Problem.
This transition from direct to indirect control is seamless and does not need mode switching or using additional interface components, leveraging the benefits of both input methods. However, one minor issue we observed from the study was that the upward sliding, for either expanding or rewinding, can cause occlusion posed by the finger body, instead of its tip. This problem was not as visible in \wordname{} as in \chunkname{}, where the user needs to actively preview the next chunks to be merged.

Another interesting phenomenon \chunkname{} brings to the design space of input control techniques is its \emph{non-static} Control-Display gain with a discrete change pattern as a function of the sliding distance. This introduces theoretical questions to target acquisition when semantics or other discrete parameters are involved.

\subsection{Outlook for NLP-Assisted Text Selection}
We believe coarse-grained text selection and NLP-assisted text selection can go hand in hand towards enabling new ways of interacting with text. Use cases of coarse-grained text selection include quick highlighting, copy-paste, translation, and selection of structured text (e.g., code), etc. These can be done with the syntactic parsing used in our techniques. Reducing the input precision needed in text selection can help smartphone users interact with text on the go, as well as benefit low-vision user groups.

Furthermore, as language AI rapidly develops, the underlying semantic structure of text will be analyzed more accurately and concretely. Future work could investigate the use of Large Language Models to extract semantic structures from text, perhaps even in a customizable or interactive way, instead of syntactic chunking used in our work. Effectively chunking text could help users consume and process large amounts of text. For instance, chunking text in transcripts of verbose speech may help users segment content based on topics or points. Chunking the text produced by generative AI may help improve the efficiency of users reviewing and processing the abundant content generated by the models.

\subsection{Beyond Touchscreen and 2D Display}
\sysname{} is shown to be efficient and easy to learn and perform thanks to its simple up-and-down sliding gesture that doesn't require extra hardware or interface components.
Using such 1D control for text selection not only benefits touchscreen input but also can be used in other display environments.
We argue that this approach shines even more when used in environments where position-based target acquisition is challenging to achieve.
For example, we can imagine our approach being used on smart glasses with a small input device like the ring mouse, or on smartwatches where targets easily go off-screen. It could also be deployed in virtual reality environments where text is displayed as floating 3D objects in non-flat layouts, making it harder to control the cursor by position in the space. To learn about the gain and cost of \sysname{} in different display and control environments, new prototypes for such environments need to be developed and evaluated in the future.

\subsection{Limitations}
\label{sec:limitation}

One main limitation of the current \chunkname{} prototype is the choice and implementation of the chunking method.
Although we propose \sysname{} techniques to be meaning-driven, our prototype doesn't extract meanings directly. 
Instead, \chunkname{} uses syntactic units to infer semantics, which tends to be limited. 
CoreNLP constituency parser also has its limitations of only looking up constituents while providing parsing results regardless of the semantic and even syntactic correctness. 
It only works relatively well when the sentence itself is grammatically regular and syntactically correct, which is not the case in many real-world scenarios. 
While the chunking methods need to be improved in the future, we worked around these limitations by ensuring the syntactic correctness of the text used in our experimental tasks.

\sysname{} is relatively sensitive to where the selection target is located. The sliding gesture needs enough vertical room to perform. Sometimes the text is near the top or the bottom, leaving limited space and may force repetitive clutching. We also noticed that when the first word of the selection target is very short, like ``I'' or ``a,'' users struggled to select it sometimes. While expert users may realize they do not need to start from the first word since the expansion is bidirectional, beginners tend not to realize this.

Although we believe our technique can be integrated with char-level selection with carets in the same way as the baseline, further studies could be conducted to assess the potential cost. One difference would be that for baseline, the position where the finger lifts after word-level selection is close to the caret position. Whereas in our case, the finger is further away due to the indirect control, potentially costing extra time for the finger to travel back to the caret position.

\section{Conclusion and Future Work}

Our work contributes a novel concept for text selection as a gradual expansion and contraction process of semantic text objects and adopts a 1D semi-direct input method to control it. This conceptual shift benefits from the increasing capability of language AI in understanding the semantic structure of text. Our work makes the first step to facilitate semantic manipulation of text by supporting coarse-grained text selection, which is a fundamental operation for most higher-level interactions with text. Our study showed that the \sysname{} approach significantly improved the state-of-the-art for coarse-grained selection by 20\%, and attributed most of the performance gain to the simple 1D semi-direct control. \sysname{} can be used as a standalone technique in coarse-grained text selection scenarios, or as a complementary method used in combination with the carets for character-level adjustment. While \wordname{} outperforms in most scenarios, \chunkname{} was shown to serve better for selecting semantically meaningful units.


We plan to improve the \chunkname{} technique for future work with enhanced and optimized chunking methods given different text target types and selection tasks.
We will try to provide ways to customize the sensitivity of distance control with empirical usage data.
\modif{Different ways of combining \wordname{} and \chunkname{} to leverage the advantages of both for varied selection tasks can also be prototyped. For example, sliding with one finger for \wordname{} and with two fingers to activate \chunkname{}.}
We will also explore the mechanism in other display environments where reducing the need for input precision is beneficial, such as tiny smartwatch screens and virtual environments.

\begin{acks}
This research was supported by the Hong Kong Research Grants Council - ECS scheme under project number CityU 21209419. We thank Shumin Zhai from Google LLC for the insightful discussions. We thank our reviewers for their constructive feedback.
\end{acks}

\bibliographystyle{ACM-Reference-Format}
\bibliography{ref}


\begin{thebibliography}{32}


\ifx \showCODEN    \undefined \def \showCODEN     #1{\unskip}     \fi
\ifx \showDOI      \undefined \def \showDOI       #1{#1}\fi
\ifx \showISBNx    \undefined \def \showISBNx     #1{\unskip}     \fi
\ifx \showISBNxiii \undefined \def \showISBNxiii  #1{\unskip}     \fi
\ifx \showISSN     \undefined \def \showISSN      #1{\unskip}     \fi
\ifx \showLCCN     \undefined \def \showLCCN      #1{\unskip}     \fi
\ifx \shownote     \undefined \def \shownote      #1{#1}          \fi
\ifx \showarticletitle \undefined \def \showarticletitle #1{#1}   \fi
\ifx \showURL      \undefined \def \showURL       {\relax}        \fi
\providecommand\bibfield[2]{#2}
\providecommand\bibinfo[2]{#2}
\providecommand\natexlab[1]{#1}
\providecommand\showeprint[2][]{arXiv:#2}

\bibitem[Ando et~al\mbox{.}(2018)]%
        {Ando2018}
\bibfield{author}{\bibinfo{person}{Toshiyuki Ando}, \bibinfo{person}{Toshiya
  Isomoto}, \bibinfo{person}{Buntarou Shizuki}, {and} \bibinfo{person}{Shin
  Takahashi}.} \bibinfo{year}{2018}\natexlab{}.
\newblock \showarticletitle{Press \& Tilt: One-Handed Text Selection and
  Command Execution on Smartphone}. In \bibinfo{booktitle}{\emph{Proceedings of
  the 30th Australian Conference on Computer-Human Interaction}} (Melbourne,
  Australia) \emph{(\bibinfo{series}{OzCHI '18})}.
  \bibinfo{publisher}{Association for Computing Machinery},
  \bibinfo{address}{New York, NY, USA}, \bibinfo{pages}{401–405}.
\newblock
\showISBNx{9781450361880}
\urldef\tempurl%
\url{https://doi.org/10.1145/3292147.3292178}
\showDOI{\tempurl}


\bibitem[Ando et~al\mbox{.}(2019)]%
        {Ando2019}
\bibfield{author}{\bibinfo{person}{Toshiyuki Ando}, \bibinfo{person}{Toshiya
  Isomoto}, \bibinfo{person}{Buntarou Shizuki}, {and} \bibinfo{person}{Shin
  Takahashi}.} \bibinfo{year}{2019}\natexlab{}.
\newblock \showarticletitle{One-Handed Rapid Text Selection and Command
  Execution Method for Smartphones}. In \bibinfo{booktitle}{\emph{Extended
  Abstracts of the 2019 CHI Conference on Human Factors in Computing Systems}}
  (Glasgow, Scotland Uk) \emph{(\bibinfo{series}{CHI EA '19})}.
  \bibinfo{publisher}{Association for Computing Machinery},
  \bibinfo{address}{New York, NY, USA}, \bibinfo{pages}{1–6}.
\newblock
\showISBNx{9781450359719}
\urldef\tempurl%
\url{https://doi.org/10.1145/3290607.3312850}
\showDOI{\tempurl}


\bibitem[Biedert et~al\mbox{.}(2010)]%
        {Biedert2010}
\bibfield{author}{\bibinfo{person}{Ralf Biedert}, \bibinfo{person}{Georg
  Buscher}, \bibinfo{person}{Sven Schwarz}, \bibinfo{person}{J\"{o}rn Hees},
  {and} \bibinfo{person}{Andreas Dengel}.} \bibinfo{year}{2010}\natexlab{}.
\newblock \bibinfo{booktitle}{\emph{Text 2.0}}.
\newblock \bibinfo{publisher}{Association for Computing Machinery},
  \bibinfo{address}{New York, NY, USA}, \bibinfo{pages}{4003–4008}.
\newblock
\showISBNx{9781605589305}
\urldef\tempurl%
\url{https://doi.org/10.1145/1753846.1754093}
\showURL{%
\tempurl}


\bibitem[Bier et~al\mbox{.}(2006)]%
        {bier2006entity}
\bibfield{author}{\bibinfo{person}{Eric~A. Bier}, \bibinfo{person}{Edward~W.
  Ishak}, {and} \bibinfo{person}{Ed Chi}.} \bibinfo{year}{2006}\natexlab{}.
\newblock \bibinfo{booktitle}{\emph{<i>Entity Quick Click</i>: Rapid Text
  Copying Based on Automatic Entity Extraction}}.
\newblock \bibinfo{publisher}{Association for Computing Machinery},
  \bibinfo{address}{New York, NY, USA}, \bibinfo{pages}{562–567}.
\newblock
\showISBNx{1595932984}
\urldef\tempurl%
\url{https://doi.org/10.1145/1125451.1125570}
\showURL{%
\tempurl}


\bibitem[Bort(2019)]%
        {bort-2019}
\bibfield{author}{\bibinfo{person}{Gerardo Bort}.}
  \bibinfo{year}{2019}\natexlab{}.
\newblock \bibinfo{title}{{CoreNLP for NodeJS}}.
\newblock
\newblock
\urldef\tempurl%
\url{https://github.com/gerardobort/node-corenlp}
\showURL{%
\tempurl}


\bibitem[Chang et~al\mbox{.}(2016)]%
        {chang2016supporting}
\bibfield{author}{\bibinfo{person}{Joseph~Chee Chang}, \bibinfo{person}{Nathan
  Hahn}, {and} \bibinfo{person}{Aniket Kittur}.}
  \bibinfo{year}{2016}\natexlab{}.
\newblock \showarticletitle{Supporting Mobile Sensemaking Through Intentionally
  Uncertain Highlighting}. In \bibinfo{booktitle}{\emph{Proceedings of the 29th
  Annual Symposium on User Interface Software and Technology}} (Tokyo, Japan)
  \emph{(\bibinfo{series}{UIST '16})}. \bibinfo{publisher}{Association for
  Computing Machinery}, \bibinfo{address}{New York, NY, USA},
  \bibinfo{pages}{61–68}.
\newblock
\showISBNx{9781450341899}
\urldef\tempurl%
\url{https://doi.org/10.1145/2984511.2984538}
\showDOI{\tempurl}


\bibitem[Chen et~al\mbox{.}(2014)]%
        {chen2014bezelcopy}
\bibfield{author}{\bibinfo{person}{Chen Chen}, \bibinfo{person}{Simon~T.
  Perrault}, \bibinfo{person}{Shengdong Zhao}, {and} \bibinfo{person}{Wei~Tsang
  Ooi}.} \bibinfo{year}{2014}\natexlab{}.
\newblock \showarticletitle{BezelCopy: An Efficient Cross-Application
  Copy-Paste Technique for Touchscreen Smartphones}. In
  \bibinfo{booktitle}{\emph{Proceedings of the 2014 International Working
  Conference on Advanced Visual Interfaces}} (Como, Italy)
  \emph{(\bibinfo{series}{AVI '14})}. \bibinfo{publisher}{Association for
  Computing Machinery}, \bibinfo{address}{New York, NY, USA},
  \bibinfo{pages}{185–192}.
\newblock
\showISBNx{9781450327756}
\urldef\tempurl%
\url{https://doi.org/10.1145/2598153.2598162}
\showDOI{\tempurl}


\bibitem[Dahl(2021)]%
        {dahl-2021}
\bibfield{author}{\bibinfo{person}{Ryan Dahl}.}
  \bibinfo{year}{2021}\natexlab{}.
\newblock \bibinfo{title}{{Node.js}}.
\newblock
\newblock
\urldef\tempurl%
\url{https://nodejs.org/}
\showURL{%
\tempurl}


\bibitem[Darbar et~al\mbox{.}(2021)]%
        {Darbar2021}
\bibfield{author}{\bibinfo{person}{Rajkumar Darbar}, \bibinfo{person}{Arnaud
  Prouzeau}, \bibinfo{person}{Joan Odicio-Vilchez}, \bibinfo{person}{Thibault
  Lain{\'e}}, {and} \bibinfo{person}{Martin Hachet}.}
  \bibinfo{year}{2021}\natexlab{}.
\newblock \showarticletitle{Exploring Smartphone-enabled Text Selection in
  AR-HMD}. In \bibinfo{booktitle}{\emph{Proceedings of Graphics Interface
  2021}} (Virtual Event) \emph{(\bibinfo{series}{GI 2021})}.
  \bibinfo{publisher}{Canadian Information Processing Society},
  \bibinfo{pages}{117 -- 126}.
\newblock
\showISBNx{978-0-9947868-6-9}
\showISSN{0713-5424}
\urldef\tempurl%
\url{https://doi.org/10.20380/GI2021.14}
\showDOI{\tempurl}


\bibitem[Esteves et~al\mbox{.}(2022)]%
        {Esteves2022One}
\bibfield{author}{\bibinfo{person}{Augusto Esteves}, \bibinfo{person}{Elizabeth
  Bouquet}, \bibinfo{person}{Ken Pfeuffer}, {and} \bibinfo{person}{Florian
  Alt}.} \bibinfo{year}{2022}\natexlab{}.
\newblock \showarticletitle{One-Handed Input for Mobile Devices via Motion
  Matching and Orbits Controls}.
\newblock \bibinfo{journal}{\emph{Proc. ACM Interact. Mob. Wearable Ubiquitous
  Technol.}} \bibinfo{volume}{6}, \bibinfo{number}{2}, Article
  \bibinfo{articleno}{51} (\bibinfo{date}{jul} \bibinfo{year}{2022}),
  \bibinfo{numpages}{24}~pages.
\newblock
\urldef\tempurl%
\url{https://doi.org/10.1145/3534624}
\showDOI{\tempurl}


\bibitem[Fuccella et~al\mbox{.}(2013)]%
        {Fuccella2013}
\bibfield{author}{\bibinfo{person}{Vittorio Fuccella}, \bibinfo{person}{Poika
  Isokoski}, {and} \bibinfo{person}{Benoit Martin}.}
  \bibinfo{year}{2013}\natexlab{}.
\newblock \bibinfo{booktitle}{\emph{Gestures and Widgets: Performance in Text
  Editing on Multi-Touch Capable Mobile Devices}}.
\newblock \bibinfo{publisher}{Association for Computing Machinery},
  \bibinfo{pages}{2785–2794}.
\newblock
\showISBNx{9781450318990}


\bibitem[Goguey et~al\mbox{.}(2018)]%
        {Goguey2018}
\bibfield{author}{\bibinfo{person}{Alix Goguey}, \bibinfo{person}{Sylvain
  Malacria}, {and} \bibinfo{person}{Carl Gutwin}.}
  \bibinfo{year}{2018}\natexlab{}.
\newblock \bibinfo{booktitle}{\emph{Improving Discoverability and Expert
  Performance in Force-Sensitive Text Selection for Touch Devices with Mode
  Gauges}}.
\newblock \bibinfo{publisher}{Association for Computing Machinery},
  \bibinfo{address}{New York, NY, USA}, \bibinfo{pages}{1–12}.
\newblock
\showISBNx{9781450356206}
\urldef\tempurl%
\url{https://doi.org/10.1145/3173574.3174051}
\showURL{%
\tempurl}


\bibitem[{Google}(2019)]%
        {Magnifier-widget}
\bibfield{author}{\bibinfo{person}{{Google}}.} \bibinfo{year}{2019}\natexlab{}.
\newblock \bibinfo{title}{{Magnifier widget}}.
\newblock
\newblock
\urldef\tempurl%
\url{https://developer.android.com/guide/topics/text/magnifier}
\showURL{%
\tempurl}


\bibitem[Hart and Staveland(1988)]%
        {HART1988139}
\bibfield{author}{\bibinfo{person}{Sandra~G. Hart} {and}
  \bibinfo{person}{Lowell~E. Staveland}.} \bibinfo{year}{1988}\natexlab{}.
\newblock \showarticletitle{Development of NASA-TLX (Task Load Index): Results
  of Empirical and Theoretical Research}.
\newblock In \bibinfo{booktitle}{\emph{Human Mental Workload}},
  \bibfield{editor}{\bibinfo{person}{Peter~A. Hancock} {and}
  \bibinfo{person}{Najmedin Meshkati}} (Eds.). \bibinfo{series}{Advances in
  Psychology}, Vol.~\bibinfo{volume}{52}. \bibinfo{publisher}{North-Holland},
  \bibinfo{pages}{139--183}.
\newblock
\showISSN{0166-4115}
\urldef\tempurl%
\url{https://doi.org/10.1016/S0166-4115(08)62386-9}
\showDOI{\tempurl}


\bibitem[Hinckley(2007)]%
        {hinckley2007input}
\bibfield{author}{\bibinfo{person}{Ken Hinckley}.}
  \bibinfo{year}{2007}\natexlab{}.
\newblock \showarticletitle{Input technologies and techniques}.
\newblock In \bibinfo{booktitle}{\emph{The human-computer interaction
  handbook}}. \bibinfo{publisher}{CRC Press}, \bibinfo{pages}{187--202}.
\newblock


\bibitem[Hinckley et~al\mbox{.}(2012)]%
        {hinckley2012informal}
\bibfield{author}{\bibinfo{person}{Ken Hinckley}, \bibinfo{person}{Xiaojun Bi},
  \bibinfo{person}{Michel Pahud}, {and} \bibinfo{person}{Bill Buxton}.}
  \bibinfo{year}{2012}\natexlab{}.
\newblock \bibinfo{booktitle}{\emph{Informal Information Gathering Techniques
  for Active Reading}}.
\newblock \bibinfo{publisher}{Association for Computing Machinery},
  \bibinfo{address}{New York, NY, USA}, \bibinfo{pages}{1893–1896}.
\newblock
\showISBNx{9781450310154}
\urldef\tempurl%
\url{https://doi.org/10.1145/2207676.2208327}
\showURL{%
\tempurl}


\bibitem[Lai et~al\mbox{.}(2019)]%
        {lai2019shortcut}
\bibfield{author}{\bibinfo{person}{Jianwei Lai}, \bibinfo{person}{Navid
  Rajabi}, {and} \bibinfo{person}{Elahe Javadi}.}
  \bibinfo{year}{2019}\natexlab{}.
\newblock \showarticletitle{A Shortcut for Caret Positioning on Touch-Screen
  Phones}. In \bibinfo{booktitle}{\emph{Proceedings of the 21st International
  Conference on Human-Computer Interaction with Mobile Devices and Services}}
  (Taipei, Taiwan) \emph{(\bibinfo{series}{MobileHCI '19})}.
  \bibinfo{publisher}{Association for Computing Machinery},
  \bibinfo{address}{New York, NY, USA}, Article \bibinfo{articleno}{35},
  \bibinfo{numpages}{6}~pages.
\newblock
\showISBNx{9781450368254}
\urldef\tempurl%
\url{https://doi.org/10.1145/3338286.3340146}
\showDOI{\tempurl}


\bibitem[LEE et~al\mbox{.}(2020)]%
        {LEE2020}
\bibfield{author}{\bibinfo{person}{Lik-Hang LEE}, \bibinfo{person}{Yiming ZHU},
  \bibinfo{person}{Yui-Pan YAU}, \bibinfo{person}{Tristan BRAUD},
  \bibinfo{person}{Xiang SU}, {and} \bibinfo{person}{Pan Hui}.}
  \bibinfo{year}{2020}\natexlab{}.
\newblock \showarticletitle{One-thumb Text Acquisition on Force-assisted
  Miniature Interfaces for Mobile Headsets}. In \bibinfo{booktitle}{\emph{2020
  IEEE International Conference on Pervasive Computing and Communications
  (PerCom)}}. \bibinfo{pages}{1--10}.
\newblock
\urldef\tempurl%
\url{https://doi.org/10.1109/PerCom45495.2020.9127378}
\showDOI{\tempurl}


\bibitem[Lee et~al\mbox{.}(2021)]%
        {Lee2021}
\bibfield{author}{\bibinfo{person}{Lik~Hang Lee}, \bibinfo{person}{Yiming Zhu},
  \bibinfo{person}{Yui-Pan Yau}, \bibinfo{person}{Pan Hui}, {and}
  \bibinfo{person}{Susanna Pirttikangas}.} \bibinfo{year}{2021}\natexlab{}.
\newblock \showarticletitle{Press-n-Paste: Copy-and-Paste Operations with
  Pressure-Sensitive Caret Navigation for Miniaturized Surface in Mobile
  Augmented Reality}.
\newblock  \bibinfo{volume}{5}, \bibinfo{number}{EICS}, Article
  \bibinfo{articleno}{199} (\bibinfo{date}{May} \bibinfo{year}{2021}),
  \bibinfo{numpages}{29}~pages.
\newblock
\urldef\tempurl%
\url{https://doi.org/10.1145/3457146}
\showDOI{\tempurl}


\bibitem[Li et~al\mbox{.}(2020)]%
        {li2020swap}
\bibfield{author}{\bibinfo{person}{Yang Li}, \bibinfo{person}{Sayan Sarcar},
  \bibinfo{person}{Sunjun Kim}, {and} \bibinfo{person}{Xiangshi Ren}.}
  \bibinfo{year}{2020}\natexlab{}.
\newblock \bibinfo{booktitle}{\emph{Swap: A Replacement-Based Text Revision
  Technique for Mobile Devices}}.
\newblock \bibinfo{publisher}{Association for Computing Machinery},
  \bibinfo{address}{New York, NY, USA}, \bibinfo{pages}{1–12}.
\newblock
\showISBNx{9781450367080}
\urldef\tempurl%
\url{https://doi.org/10.1145/3313831.3376217}
\showURL{%
\tempurl}


\bibitem[Manning et~al\mbox{.}(2014)]%
        {manning2014stanford}
\bibfield{author}{\bibinfo{person}{Christopher~D Manning},
  \bibinfo{person}{Mihai Surdeanu}, \bibinfo{person}{John Bauer},
  \bibinfo{person}{Jenny~Rose Finkel}, \bibinfo{person}{Steven Bethard}, {and}
  \bibinfo{person}{David McClosky}.} \bibinfo{year}{2014}\natexlab{}.
\newblock \showarticletitle{The Stanford CoreNLP natural language processing
  toolkit}. In \bibinfo{booktitle}{\emph{Proceedings of 52nd annual meeting of
  the association for computational linguistics: system demonstrations}}.
  \bibinfo{pages}{55--60}.
\newblock


\bibitem[Miura and Saisho(2014)]%
        {MIURA20141644}
\bibfield{author}{\bibinfo{person}{Motoki Miura} {and} \bibinfo{person}{Kenji
  Saisho}.} \bibinfo{year}{2014}\natexlab{}.
\newblock \showarticletitle{A Text Selection Technique Using Word Snapping}.
\newblock \bibinfo{journal}{\emph{Procedia Computer Science}}
  \bibinfo{volume}{35} (\bibinfo{year}{2014}), \bibinfo{pages}{1644--1651}.
\newblock
\showISSN{1877-0509}
\urldef\tempurl%
\url{https://doi.org/10.1016/j.procs.2014.08.257}
\showDOI{\tempurl}
\newblock
\shownote{Knowledge-Based and Intelligent Information \& Engineering Systems
  18th Annual Conference, KES-2014 Gdynia, Poland, September 2014 Proceedings}.


\bibitem[Pantel and Gamon(2014)]%
        {pantel-etal-2014-smart}
\bibfield{author}{\bibinfo{person}{Patrick Pantel} {and} \bibinfo{person}{Ariel
  Gamon, Michael~andFuxman}.} \bibinfo{year}{2014}\natexlab{}.
\newblock \showarticletitle{Smart Selection}. In
  \bibinfo{booktitle}{\emph{Proceedings of the 52nd Annual Meeting of the
  Association for Computational Linguistics (Volume 1: Long Papers)}}.
  \bibinfo{publisher}{Association for Computational Linguistics},
  \bibinfo{address}{Baltimore, Maryland}, \bibinfo{pages}{1524--1533}.
\newblock
\urldef\tempurl%
\url{https://doi.org/10.3115/v1/P14-1143}
\showDOI{\tempurl}


\bibitem[Pfeuffer et~al\mbox{.}(2015)]%
        {Pfeuffer2015gaze}
\bibfield{author}{\bibinfo{person}{Ken Pfeuffer}, \bibinfo{person}{Jason
  Alexander}, \bibinfo{person}{Ming~Ki Chong}, \bibinfo{person}{Yanxia Zhang},
  {and} \bibinfo{person}{Hans Gellersen}.} \bibinfo{year}{2015}\natexlab{}.
\newblock \showarticletitle{Gaze-Shifting: Direct-Indirect Input with Pen and
  Touch Modulated by Gaze}. In \bibinfo{booktitle}{\emph{Proceedings of the
  28th Annual ACM Symposium on User Interface Software \& Technology}}
  (Charlotte, NC, USA) \emph{(\bibinfo{series}{UIST '15})}.
  \bibinfo{publisher}{Association for Computing Machinery},
  \bibinfo{address}{New York, NY, USA}, \bibinfo{pages}{373–383}.
\newblock
\showISBNx{9781450337793}
\urldef\tempurl%
\url{https://doi.org/10.1145/2807442.2807460}
\showDOI{\tempurl}


\bibitem[Reyna and Brainerd(1995)]%
        {REYNA19951}
\bibfield{author}{\bibinfo{person}{V.F. Reyna} {and} \bibinfo{person}{C.J.
  Brainerd}.} \bibinfo{year}{1995}\natexlab{}.
\newblock \showarticletitle{Fuzzy-trace theory: An interim synthesis}.
\newblock \bibinfo{journal}{\emph{Learning and Individual Differences}}
  \bibinfo{volume}{7}, \bibinfo{number}{1} (\bibinfo{year}{1995}),
  \bibinfo{pages}{1--75}.
\newblock
\showISSN{1041-6080}
\urldef\tempurl%
\url{https://doi.org/10.1016/1041-6080(95)90031-4}
\showDOI{\tempurl}
\newblock
\shownote{Special Issue: Fuzzy-Trace Theory}.


\bibitem[Rivu et~al\mbox{.}(2020)]%
        {Rivu2020}
\bibfield{author}{\bibinfo{person}{Radiah Rivu}, \bibinfo{person}{Yasmeen
  Abdrabou}, \bibinfo{person}{Ken Pfeuffer}, \bibinfo{person}{Mariam Hassib},
  {and} \bibinfo{person}{Florian Alt}.} \bibinfo{year}{2020}\natexlab{}.
\newblock \showarticletitle{Gaze'N'Touch: Enhancing Text Selection on Mobile
  Devices Using Gaze}. In \bibinfo{booktitle}{\emph{Extended Abstracts of the
  2020 CHI Conference on Human Factors in Computing Systems}} (Honolulu, HI,
  USA) \emph{(\bibinfo{series}{CHI EA '20})}. \bibinfo{publisher}{Association
  for Computing Machinery}, \bibinfo{address}{New York, NY, USA},
  \bibinfo{pages}{1–8}.
\newblock
\showISBNx{9781450368193}
\urldef\tempurl%
\url{https://doi.org/10.1145/3334480.3382802}
\showDOI{\tempurl}


\bibitem[Roudaut et~al\mbox{.}(2008)]%
        {Roudaut2008}
\bibfield{author}{\bibinfo{person}{Anne Roudaut}, \bibinfo{person}{St\'{e}phane
  Huot}, {and} \bibinfo{person}{Eric Lecolinet}.}
  \bibinfo{year}{2008}\natexlab{}.
\newblock \showarticletitle{TapTap and MagStick: Improving One-Handed Target
  Acquisition on Small Touch-Screens}. In \bibinfo{booktitle}{\emph{Proceedings
  of the Working Conference on Advanced Visual Interfaces}} (Napoli, Italy)
  \emph{(\bibinfo{series}{AVI '08})}. \bibinfo{publisher}{Association for
  Computing Machinery}, \bibinfo{address}{New York, NY, USA},
  \bibinfo{pages}{146–153}.
\newblock
\showISBNx{9781605581415}
\urldef\tempurl%
\url{https://doi.org/10.1145/1385569.1385594}
\showDOI{\tempurl}


\bibitem[Siek et~al\mbox{.}(2005)]%
        {siek2005fat}
\bibfield{author}{\bibinfo{person}{Katie~A Siek}, \bibinfo{person}{Yvonne
  Rogers}, {and} \bibinfo{person}{Kay~H Connelly}.}
  \bibinfo{year}{2005}\natexlab{}.
\newblock \showarticletitle{Fat finger worries: how older and younger users
  physically interact with PDAs}. In \bibinfo{booktitle}{\emph{IFIP Conference
  on Human-Computer Interaction}}. Springer, \bibinfo{pages}{267--280}.
\newblock


\bibitem[Suzuki et~al\mbox{.}(2016)]%
        {suzuki2016fix}
\bibfield{author}{\bibinfo{person}{Kenji Suzuki}, \bibinfo{person}{Kazumasa
  Okabe}, \bibinfo{person}{Ryuuki Sakamoto}, {and} \bibinfo{person}{Daisuke
  Sakamoto}.} \bibinfo{year}{2016}\natexlab{}.
\newblock \showarticletitle{Fix and Slide: Caret Navigation with Movable
  Background}. In \bibinfo{booktitle}{\emph{Proceedings of the 18th
  International Conference on Human-Computer Interaction with Mobile Devices
  and Services}} (Florence, Italy) \emph{(\bibinfo{series}{MobileHCI '16})}.
  \bibinfo{publisher}{Association for Computing Machinery},
  \bibinfo{address}{New York, NY, USA}, \bibinfo{pages}{478–482}.
\newblock
\showISBNx{9781450344081}
\urldef\tempurl%
\url{https://doi.org/10.1145/2935334.2935357}
\showDOI{\tempurl}


\bibitem[Zhang and Wobbrock(2019)]%
        {zhang2019gedit}
\bibfield{author}{\bibinfo{person}{Mingrui Zhang} {and}
  \bibinfo{person}{Jacob~O Wobbrock}.} \bibinfo{year}{2019}\natexlab{}.
\newblock \showarticletitle{Gedit: Keyboard gestures for mobile text editing}.
\newblock  (\bibinfo{year}{2019}).
\newblock


\bibitem[Zhao et~al\mbox{.}(2022)]%
        {10.1145/3490099.3511103}
\bibfield{author}{\bibinfo{person}{Maozheng Zhao}, \bibinfo{person}{Henry
  Huang}, \bibinfo{person}{Zhi Li}, \bibinfo{person}{Rui Liu},
  \bibinfo{person}{Wenzhe Cui}, \bibinfo{person}{Kajal Toshniwal},
  \bibinfo{person}{Ananya Goel}, \bibinfo{person}{Andrew Wang},
  \bibinfo{person}{Xia Zhao}, \bibinfo{person}{Sina Rashidian},
  \bibinfo{person}{Furqan Baig}, \bibinfo{person}{Khiem Phi},
  \bibinfo{person}{Shumin Zhai}, \bibinfo{person}{IV Ramakrishnan},
  \bibinfo{person}{Fusheng Wang}, {and} \bibinfo{person}{Xiaojun Bi}.}
  \bibinfo{year}{2022}\natexlab{}.
\newblock \showarticletitle{EyeSayCorrect: Eye Gaze and Voice Based Hands-Free
  Text Correction for Mobile Devices}. In \bibinfo{booktitle}{\emph{27th
  International Conference on Intelligent User Interfaces}} (Helsinki, Finland)
  \emph{(\bibinfo{series}{IUI '22})}. \bibinfo{publisher}{Association for
  Computing Machinery}, \bibinfo{address}{New York, NY, USA},
  \bibinfo{pages}{470–482}.
\newblock
\showISBNx{9781450391443}
\urldef\tempurl%
\url{https://doi.org/10.1145/3490099.3511103}
\showDOI{\tempurl}


\bibitem[Zhu et~al\mbox{.}(2013)]%
        {zhu2013fast}
\bibfield{author}{\bibinfo{person}{Muhua Zhu}, \bibinfo{person}{Yue Zhang},
  \bibinfo{person}{Wenliang Chen}, \bibinfo{person}{Min Zhang}, {and}
  \bibinfo{person}{Jingbo Zhu}.} \bibinfo{year}{2013}\natexlab{}.
\newblock \showarticletitle{Fast and accurate shift-reduce constituent
  parsing}. In \bibinfo{booktitle}{\emph{Proceedings of the 51st Annual Meeting
  of the Association for Computational Linguistics (Volume 1: Long Papers)}}.
  \bibinfo{pages}{434--443}.
\newblock


\end{thebibliography}

\clearpage
\onecolumn
\appendix
\section{Post hoc Test Results}
\label{app:test}
\begin{table*}[h!]
\caption{Paired post hoc tests for test completion time}
\begin{tabular}{cccccc}
\toprule
Task Type     & Technique A & Technique B & $W$ Value    & $p$ Value \\
\midrule
Word          & Baseline    & ChunkTouch  & 20     & 0.255     \\
Word          & Baseline    & WordTouch   & 27     & 0.514     \\
Word          & ChunkTouch  & WordTouch   & 32     & 0.730     \\
Phrase        & Baseline    & ChunkTouch  & 6      & 0.031     \\
Phrase        & Baseline    & WordTouch   & 9      & 0.036     \\
Phrase        & ChunkTouch  & WordTouch   & 21     & 0.264     \\
Non-phrase    & Baseline    & ChunkTouch  & 36     & 0.850     \\
Non-phrase    & Baseline    & WordTouch   & 0      & 0.013     \\
Non-phrase    & ChunkTouch  & WordTouch   & 10     & 0.044     \\
Clause        & Baseline    & ChunkTouch  & 8      & 0.033     \\
Clause        & Baseline    & WordTouch   & 5      & 0.031     \\
Clause        & ChunkTouch  & WordTouch   & 34     & 0.762     \\
Half-sentence & Baseline    & ChunkTouch  & 34     & 0.762     \\
Half-sentence & Baseline    & WordTouch   & 9      & 0.036     \\
Half-sentence & ChunkTouch  & WordTouch   & 8      & 0.033     \\
Sentence      & Baseline    & ChunkTouch  & 8      & 0.033     \\
Sentence      & Baseline    & WordTouch   & 2      & 0.013     \\
Sentence      & ChunkTouch  & WordTouch   & 29     & 0.604     \\
Two-sentences & Baseline    & ChunkTouch  & 8      & 0.033     \\
Two-sentences & Baseline    & WordTouch   & 33     & 0.762     \\
Two-sentences & ChunkTouch  & WordTouch   & 20     & 0.255     \\
Word-to-end   & Baseline    & ChunkTouch  & 2      & 0.013     \\
Word-to-end   & Baseline    & WordTouch   & 18     & 0.212     \\
Word-to-end   & ChunkTouch  & WordTouch   & 6      & 0.031     \\
Whole-text    & Baseline    & ChunkTouch  & 24     & 0.378     \\
Whole-text    & Baseline    & WordTouch   & 30     & 0.636     \\
Whole-text    & ChunkTouch  & WordTouch   & 21     & 0.264     \\ 
\bottomrule
\end{tabular}
\label{table:1}
\end{table*}

\end{document}